\documentclass[useAMS,usenatbib]{mn2e}
 \usepackage{times}
 \usepackage{graphicx, latexsym, amssymb, amsmath, amscd, psfrag}
 \usepackage{epsfig}
\usepackage{color}
\usepackage{journals}
 \def\eg{{e.g.}}
 \def\ie{{i.e.}}
 \def\ssf{{$\log$ SFR/$M^*$}}
 \def\hd{{H$\delta$ }}
 \def\ehd{{EW(H$\delta$)}}
 \def\rv{{$R_{\rm v}$ }}

 \def\simgt{\hbox{\rlap{\raise 0.425ex\hbox{$>$}}\lower 0.65ex\hbox{$\sim$}}}
 \def\simlt{\hbox{\rlap{\raise 0.425ex\hbox{$<$}}\lower 0.65ex\hbox{$\sim$}}}

\definecolor{grey}{rgb}{0.5,0.6,0.7}

 \title[Velocity modulation of galaxy properties]
{The velocity modulation of galaxy properties in and near clusters: 
 quantifying the decrease in star formation in backsplash galaxies}

 \author[Mahajan, Mamon and Raychaudhury]{Smriti Mahajan$^{1}$\thanks{E-mail: sm@star.sr.bham.ac.uk}, 
 Gary A. Mamon$^{2,3}$ and Somak Raychaudhury$^{1}$ \\  
 1: School of Physics and Astronomy, University of Birmingham, Birmingham B15~2TT,
 United Kingdom \\
 2: Institut d'Astrophysique de Paris (IAP), 98-bis Blvd Arago,
 F-75014 Paris, France\\
3: Astrophysics \& BIPAC, Department of Physics, University of Oxford, Keble
Rd, Oxford OX1 3RH, United Kingdom
}

\begin{document}

 \date{}

 \pagerange{\pageref{firstpage}--\pageref{lastpage}} \pubyear{2010}

 \maketitle

 \label{firstpage}
 \begin{abstract}
   Ongoing/recent star formation in galaxies is
   known to increase with increasing projected distance from the
   centre of a cluster out to several times its virial radius ($R_{\rm
     v}$). Using a complete sample ($M_{r}\!\leq\!-20.5$,
   $0.02\!\leq\!z\leq\!0.15$) of galaxies in and around 268 clusters from 
the SDSS DR4, we investigate how, at a
   given projected radius from the cluster centre, the stellar mass
   and star formation properties of a galaxy depend on its absolute
   line-of-sight  velocity in the cluster rest frame, $|v_{\rm
     LOS}|$.  We find that for projected radii $R\!<\!0.5\,R_{\rm v}$,
   the fraction of high mass non-BCG galaxies increases towards the
   centre for low $|v_{\rm LOS}|$, which may be the consequence of the
   faster orbital decay of massive galaxies by dynamical friction. At
   a given projected radius, the fraction of Galaxies with Ongoing or
   Recent ($<$1--3 Gyr) Efficient Star Formation (GORES, with
   \ehd\,$>\!2$\,\AA~\& D$_n$4000\,$>\!1.5$) is slightly but
   significantly lower for low $|v_{\rm LOS}|$ galaxies than for their
   high velocity counterparts.  We study these observational trends
   with the help of a dark matter (DM) cosmological simulation. We
   classify DM particles as virial, infall, and backsplash according
   to their present positions in ($r, v_r$) radial phase space and
   measure the frequencies of each class in cells of ($R,|v_{\rm
     LOS}|$) projected phase space. As expected, the virial class
   dominates at projected radii $R\!<\!R_{\rm v}$, while the infall
   particles dominate outside, especially at high $|v_{\rm
     LOS}|$. However, the backsplash particles account for at least
   one-third (half) of all particles at projected radii slightly
   greater than the virial radius and $|v_{\rm LOS}|\!<\!\sigma_v$
   ($|v_{\rm LOS}|\!\ll\!\sigma_v$).  The deprojection of the GORES
   fraction leads to a saturated linear increase with radius. We fit
   simple models of the fraction of GORES as a function of class only
   or class and distance to the cluster centre (as in our deprojected
   fraction).  While GORES account for $18\pm1\%$ of all galaxies within the
   virial cylinder, in our best-fit model, they account for $13\pm1\%$ of galaxies
   within the virial sphere, $11\pm1\%$ of
   the virial population, $34\pm1\%$ of the distant (for projected
   radii $R<2\,R_{\rm v}$) infall population and $19\pm4\%$ of the
   backsplash galaxies. Also, $44\pm2\%$ of the GORES within the virial cylinder are
   outside the virial sphere. These fractions are very robust to the
   precise good-fitting model and to our scheme for assigning
   simulation particle classes according to their positions in radial
   phase space (except for two of our models, where the fraction of GORES
   reaches $27\pm4\%$).  Given the 1--3 Gyr lookback time of our GORES 
   indicators, these results suggest that star formation in a galaxy is
   almost completely quenched in a single passage through the cluster.
 \end{abstract}


 \section{Introduction}
\label{intro} 
 The wide variety in star formation properties of galaxies is an
 inevitable consequence of the varied environments they inhabit.
 Physical properties of galaxies such as morphology \citep{dressler80}, colour
 \citep[\eg][]{balogha04}, luminosity (e.g. \citealp*{ABM98}),
 individual spectral indices used as star formation rate (SFR) tracers, such as the
 equivalent width (EW) of [OII] or H$\alpha$
 (\eg\ \citealp*{baloghb04,Haines+06}), are known to vary systematically with
 environment. As they leave their
 sparsely populated base in the \emph{field} for the high density regions of
 galaxy clusters,  galaxies see their dominant stellar population become older,
 emission lines disappear from their spectra,
 complex spirals are transformed into more uniform spheroids, and the SFR
 decreases by 1-2 orders of magnitude.

It is thus generally believed that the cluster environment is hostile to star
formation. For example, the denser the environment, the stronger will be the
stripping of interstellar gas either by the ram pressure of the intra-cluster
gas \citep{gunn72} or by the tidal field of the cluster \citep{merritt83},
leading to starvation for subsequent star formation \citep*{larson80}.
So a galaxy that entered the cluster virial radius $\sim$\,1 Gyr ago, and has since
 crossed the core of the cluster, should have lost its gas via either of
 these two effects during its passage through the cluster core
(see for instance \citealp{vollmer09} for the impact of ram-pressure stripping in the Virgo
 cluster). Moreover, in the intermediate density environments
 on the cluster outskirts, mechanisms like galaxy-galaxy interactions
 \citep[\eg][]{moore96} seem to dominate the process of galaxy evolution
(\eg\ \citealp{porter07}; \citealp{porter08}; \citealp*{mahajan11}). Thus star-forming
 \emph{infalling} galaxies are expected to have distinctive properties
 compared to the passively evolving \emph{virialised} population
 inside the cluster.

However, one can define an intermediate population of galaxies that have
travelled through the cluster core (perhaps more than once) 
and have not yet had time to mix with the
virialised population.  
 These \emph{backsplash}\footnote{The term \emph{backsplash galaxies} was first
   coined by \citet*{gill05}.} galaxies would have felt the first effects of the
 cluster environment, but may not have evolved all the way to attain the
 passive properties of the virialised population.
In their simulations of six clusters,
 \citet*{balogh00} find that 54\%\,$\pm$\,20\% of dark matter (DM) particles at
 $1\!-\!2\,R_{200}$ from the cluster centre have been inside the virial
 radius ($R_{200}$) at some earlier time. \citet{gill05} estimate
 similar fractions, considering a virial radius equivalent to
 $R_{100}$, and further add that 90\% of such particles go as close to
 the cluster centre as 50\% of the virial radius. \citet{mamon04} used both
 analytical methods and cosmological simulations to conclude that DM
 particles travel through the core of the cluster out to
 1--2.5$\,R_{100}$ on the other side. 

 The backsplash galaxies may contribute to the observation that the
 radial trends of star formation diagnostics extend to at least twice
 the virial radius \citep{balogh00} from cluster centres. Hence, the
 intermediate environments found in inter-cluster filaments and on
 the outskirts of galaxy clusters play a key role in defining the
 evolutionary history of galaxies
 \citep{rines03,gill05,porter07,porter08,mahajan10,mahajan11}.

 Most results indicate that the properties of galaxies evolve as they
 encounter denser environments, \ie\ progress from the field towards
 the cluster.  This could be due to the prevalence of smaller clusters
 (galaxy groups) as one approaches rich clusters, as the tidal field
 and possibly some mild form of ram pressure stripping might play a
 similar role, as they do in more massive clusters, to deplete the gas
 reservoirs of galaxies.  Alternatively, it is possible that the gas
 reserves of galaxies may be protected by the pressure of the surrounding intra-group
 medium (IGM) from being tidally perturbed, preventing the star
 formation in the galaxy from being quenched, even after spending
 $>\!1$ Gyr in crossing the cluster.  In such a case, an increase in
 tides generated during the group--cluster merger may instead result in an
 enhancement of the SFR of group galaxies 
 \citep{bekki99}. Although such a group pre-processing scenario is yet to be confirmed,
 some evidence in its favour is found in recent observational
 studies \citep[\eg][]{oemler09}.

The main difference between the infalling and backsplash galaxies is that the
former have larger apocentres, hence larger radial (3D) velocities.
Hence, one expects that the backsplash galaxies will have lower absolute
line-of-sight (hereafter, LOS) velocities\footnote{Our absolute LOS velocities
  are in the cluster frame.} than the infalling galaxies. One is therefore
motivated to study the modulation of the radial trends of galaxy properties,
 with absolute LOS velocity. 

 In the central regions of clusters, the relative velocities of
 galaxies have been found to be correlated with their luminosity
 \citep[\eg][]{chincarini77,struble79,bothun90,biviano92}, morphology
 \citep[][ among others]{devaucouleurs61,sandage76,moss77,helou79,sodre89,biviano92,girardi03},
 and more recently spectral type \citep{biviano97,pdh06}. The
 early-type, massive and passive galaxies are found to have lower
 velocity dispersion relative to their late-type,
 star-forming counterparts. It has been shown that this
 velocity segregation of mass (or luminosity) in cluster galaxies is neither
 induced by morphological segregation nor limited to cD galaxies 
 \citep{biviano92}.

There have been some efforts to study the velocity segregation in
 galaxy properties on the outskirts of galaxy clusters \citep[\eg][]{mohr96,rines05,pdh06}.
In particular, \cite{rines05} found no significant
 difference in the velocity distribution of the emission-line and absorption-line
 galaxies with projected radii between 1 and $2\,R_{200}$ in a sample of eight
 nearby x-ray bright clusters.
However, emission-line galaxies outside the virial radius appear to have a
significantly larger velocity dispersion than their absorption-line
counterparts (but \citeauthor{rines05} do not provide any quantitative estimate of the
significance). Moreover, from fig.~6 of \citeauthor{rines05}, 
the mean velocity dispersion profile of emission-line galaxies appears
significantly steeper than that of non-emission-line galaxies (this trend was
not noted by the authors).
\citeauthor{rines05} concluded that star-forming galaxies in the infall regions are not
field interlopers, which lead them to reject a simple model where
the truncation of star formation at low projected radius was solely caused by
the efficient quenching of star formation in (backsplash) galaxies that have crossed
through the cluster.
  
 In this paper, using the very large Sloan Digital Sky Survey (SDSS), we have
 collected sufficient data around hundreds of clusters to analyse the
 differences in the distribution in \emph{projected phase space}
 (projected radius $R$ and absolute LOS velocity $|v_{\rm LOS}|$)
of star-forming and
 passive galaxies at much finer resolution than previously performed.
We look for and find a significant velocity modulation of the variations in
stellar mass and star-formation properties with projected radius. We also build
 simple dynamical models involving the virial, infalling and
 backsplash populations of galaxies to explain our results, calibrating our models and the
 projection effects using a cosmological simulation. 
 
   The organisation of this paper is as follows: in the following section we describe the
   observational dataset and briefly describe our star formation diagnostics.
In \S\ref{analysis} we present the analysis of the SDSS galaxies around
clusters, focusing on the velocity modulation of radial trends in stellar
mass and star formation efficiency.
In \S\ref{models} we model our results using a cosmological simulation
described in \S\ref{simulations} to classify particles into virial, infall
and backsplash populations (\S\ref{classif}). In \S\ref{model12}, we fit
simple models of the variation with physical radius of the fraction of
\emph{Galaxies with Ongoing or Recent Efficient Star formation} (hereafter,
GORES) for each of the three populations 
to the
observed fraction of GORES as a function of projected
radius and LOS velocity.
   We discuss and summarise our results and their implications in
   \S\ref{discussion}.
 Throughout this paper we assume a concordance cosmology with
   $\Omega_{\rm m}\!=\!0.3$, $\Omega_\Lambda\!=\!0.7$, and $h\!=\!0.7$.

 \section{Data}
 \label{data}

 \subsection{The cluster catalogue}
 
 We selected our galaxy clusters from the `group' catalogue compiled by
 \citet*{yang07} 
 from the Fourth Data Release (DR4) of the SDSS.
 Their iterative algorithm identifies first order groups using a
 Friends-of-Friends (FoF) algorithm in redshift space, with a small
 linking length in projected space and a large one along the direction
 of the LOS. This algorithm thus selects potential
 groups at a high surface density threshold, so that groups embedded
 within the virial radii of more massive ones will be extracted as two
 (or more) separate entities.  All galaxies left out by this grouping
 method are treated as additional potential groups (of single
 galaxies).

The raw luminosities and (luminosity-weighted) centres of the clusters are measured
and then the luminosities (corrected for incompleteness), masses
(assuming a given mass-to-light ratio) and velocity dispersions (from
a mass / velocity dispersion relation) are derived. The radii
$R_{180}$, \ie~of spheres that are 180 times denser than the critical
density of the Universe are derived from the masses.
The distribution of galaxies around each group centre in projected
phase space are examined to determine group membership, based upon the
assumption of a projected \citeauthor*{NFW96} (\citeyear{NFW96},
hereafter NFW) surface density profile and a Gaussian distribution of
LOS velocities with dispersion independent of projected radius. The
algorithm is iterated with these new group assignments, where the
mass-to-light ratio used for estimating the new mass is a function of
luminosity derived from the previous iteration.  The iteration
continues until the membership converges.  \citet{yang07} have tested
their algorithm on mock galaxy catalogues constructed in real space and
converted to redshift space.
  
In the present work,
 we define the galaxy clusters with the following criteria:
\begin{enumerate}
\item redshift range: $0.02\leq z\leq 0.12$;
\item membership: at least 15 galaxies;
\item minimum halo mass (within $R_{180}$): $10^{14}\rm M_\odot$.
\end{enumerate}
All of our chosen clusters have at least 12 galaxies
($M_{r}\!\leq\!-20.5$) with SDSS spectra within $R_{180}$ of each
cluster centre.  For our chosen cosmology, the virial radius, within
which the cluster may be thought to be in dynamical equilibrium, is
commonly believed to be very close to $R_{100}$ where the density is
100 times critical.
In this paper, we adopt $R_{100}$ for the
virial radius, and more precisely
\begin{equation}
R_{ v} = 1.3\,R_{180} \ ,
\label{RviroverR180}
\end{equation} 
which,
for NFW
models with concentrations between 3 and 10, matches $R_{100}$ to an accuracy
of 2\%. This gives us a sample of 268 clusters in the SDSS DR4 footprint
with median $M_{180}\!\sim\!10^{14.2} {\rm M_\odot}$,
corresponding to a median virial radius $R_{100} = 1.5\,\rm Mpc$ (using
Eq.~\ref{RviroverR180}).

The range of 1.5 dex in cluster masses should not be a concern given that the
properties of non-central (satellite) galaxies depend little on the mass of
the group or cluster they live in \citep{vandenBosch+08}.

 \subsection{Galaxy data}
 \label{galaxy}
 
 The observed photometric and spectroscopic data for galaxies
in this work are taken from the SDSS DR4 \citep{a2}. 
We use the galaxy magnitudes, $k$-corrections and the corresponding galactic extinction
values in the SDSS $g$ and $r$-bands from the New York University Value Added
Galaxy Catalogue \citep[NYU-VAGC;][]{b5}. 
We $k$-correct these
magnitudes to $z\!=\!0.1$, which is the median redshift of the sample.
We select galaxies that meet the following criteria:
\begin{enumerate}
\item redshift range $0.02 \leq z \leq 0.15$;
\item absolute magnitude: $M_r \leq -20.5$;
\item half-light angular radius: $\theta_{\rm eff} \leq 5''$;
\item projected separation with cluster: $R < 2\,R_{\rm v}$;
\item velocity relative to cluster: $\left |v_{\rm LOS}\!-\!v_{\rm
  LOS}^{\rm cluster}\right| \leq 3\,\sigma_{v}^{\rm cluster}$.
\end{enumerate}
Our luminosity limit roughly corresponds to $L\geq 0.5\,L*$ (given that 
 $M_{r,*}^{0.1}\!=\!-20.44\,+\!5\,\log\,h$, \citealp{blanton03}), 
which corresponds to the apparent magnitude
limit ($r\!\leq\!17.77$) for the SDSS spectroscopic catalogue at
$z\!=\!0.1$. Our galaxy redshifts are chosen in an interval that is slightly
larger than for the clusters, so as to avoid missing galaxies lying behind (or
in front) of the most (least) distant clusters.
The angular size criterion ensures that 
the spectrum obtained from the 3$^{\prime\prime}$ SDSS fibre
is a fair representation of the light from the galaxy, and does not
just correspond to the passive central bulge of the galaxy.

 This gives us a sample of 19,904 galaxies in or near 268 clusters covered
 out to 2 virial radii in projection and 3 times the cluster velocity dispersions along
 the LOS.
  
The cluster galaxies are compared with a sample of 21,100 field
galaxies, randomly chosen to have the same redshift and luminosity
distributions as those of the cluster galaxies. We require the field
galaxies to lie at least 10~Mpc away from the centres of all groups with
$N\!\geq\!5$ and halo mass $\geq\!10^{12.5}\,{\rm M_\odot}$ in the
\citet{yang07} catalogue (this includes clusters).
Since the galaxy properties depend on their distance from the nearest large
 structure (e.g. \citealp{mahajan11}) out to $\sim\!3 R_{200}$
($\lesssim\!6\,h_{70}^{-1}$\,Mpc; e.g. \citealp{rines03,mahajan11}), a
 linear distance of 10~Mpc (corresponding to typically $6.7\,R_{\rm v}$) 
from all groups and clusters ensures that
 our field sample is not contaminated by the effect of any groups or
 clusters.

\subsection{Star formation diagnostics}
\label{sfdiag}
In this paper, as our principal diagnostic, we use the specific star
formation rate (SFR/$M^*$) of a galaxy, as estimated from the optical
spectrum, by \citet{b04} for all galaxies belonging to the
spectroscopic catalogue of SDSS DR4. From this and a few other spectral
indices, namely the 
emission-corrected H$\delta_A$ (hereafter, H$\delta$) EW and
the 4000\,\AA~break, estimated in a related work \citep{kauff03a}, we
compare the trends seen in the current star formation activity and
star formation histories (SFHs) of different types of galaxies.

  \begin{figure*}
 \centering{
 {\rotatebox{270}{\epsfig{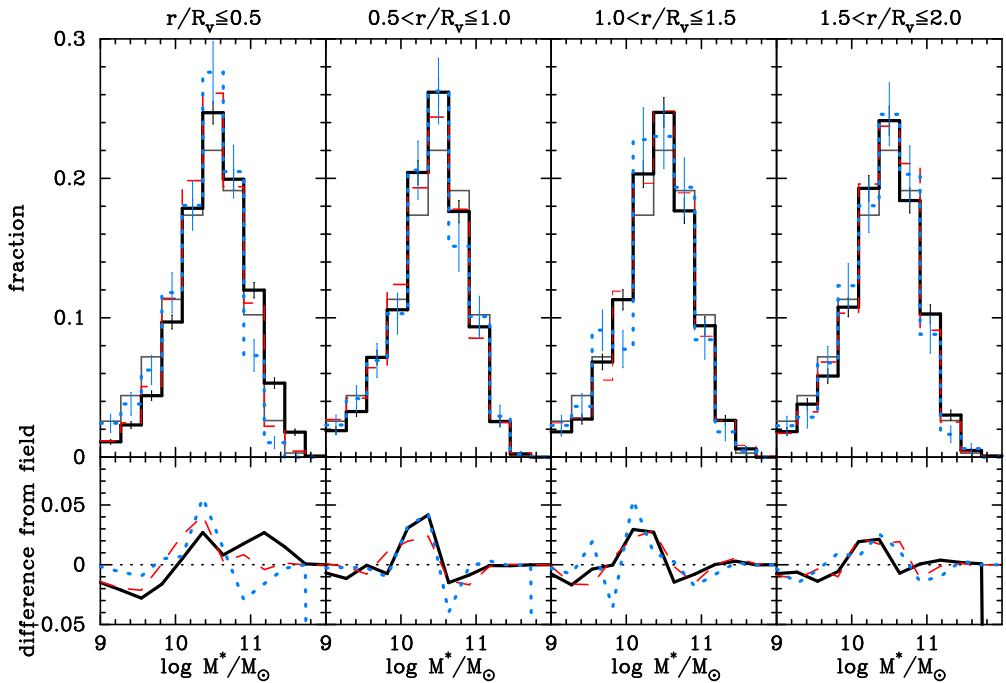}}}}
 \caption{The distribution of stellar masses for our galaxy samples, in four
   bins of projected radius in the four panels from left to right. The
 \emph{solid black}, \emph{dashed red} and \emph{dotted blue} distributions
 represent absolute LOS velocity
bins $|v_{\rm LOS}|/\sigma_{v} =$ 0--1, 1--2 and 2--3 respectively. The
  \emph{thin grey distribution} denotes the field. The \emph{bottom panels} show
  the difference between the binned cluster distributions and that of the field. 
  The \emph{error bars} are Poisson. In the flux limited SDSS,
   galaxies in the local Universe have their $M^*$ distribution centred around
   $M^{*}=3\times10^{10}$ ${\rm M_\odot}$ \citep{kauff03a}. A significant fraction of
   galaxies with $M^{*}\geq 10^{11}$ ${\rm M_\odot}$ are found only in the
   core of galaxy clusters. Interestingly, the low absolute LOS velocity galaxies in
   the cores are the ones which are more massive, while the high
   velocity galaxies mostly inhabit the low mass end. }
 \label{lgm}
 \end{figure*}

 \citeauthor{b04} have constructed a model grid of
 $\sim\!2\!\times\!10^5$ spectral synthesis models
 \citep{bruzual93,charlot01} over four main parameters spanning a
 range of $-1\!<\!\log Z/Z_\odot\!<\!0.6$ in metallicity ($Z$),
 $-4.0\!<\!\log U\!<\!-2.0$ in ionisation parameter ($U$),
 $0.01\!<\!\tau_V\!<\!4.0$ in total dust extinction ($\tau_V$) and
 $0.1\!<\!\xi\!<\!0.5$ in dust-to-metal ratio ($\xi$). Dust attenuation
 is also treated with a very sophisticated model of \citet{charlot00}, which provides
 a consistent model for ultraviolet (UV) to infrared (IR) emission.
 They then go on to derive the values for the SFR and SFR/$M^*$ using
 a Bayesian technique, considering the entire spectrum of each
 galaxy. The derived parameters (SFR and SFR/$M^*$) are then corrected
 for aperture biases using photometric colours.  Their analysis of
 line ratios is insensitive to stellar ages, SFH and the relative
 attenuation by dust in the gas clouds and the interstellar medium.
 Out of the three statistical estimates (mean, median \& mode) for the
 probability distribution function of SFR/$M^*$ derived for each
 galaxy, we used the median of the distribution in this paper, since it
 is independent of binning.

 The spectrum of a galaxy that has experienced a burst of star formation in the recent past
 (1--3 Gyr ago) is expected to show a strong absorption of \hd in the optical
 spectrum. This is because A~type stars, which significantly contribute to
 the \hd absorption, have a lifetime of $\lesssim\!1.5$\,Gyr, but a burst of
 star formation can produce enough ionising radiation to fill in the \hd
 absorption. However, as the episode of star formation ends, the \hd absorption line
 becomes more transparent. The EW(\hd) has been extensively used in the literature to
 measure the mean stellar age of galaxies, and the elapsed time since the last major burst
 \citep[\eg][]{worthey97,kauff03a}. At the other extreme is
 the 4000\,\AA~break (D$_n$4000), which is the strongest discontinuity occurring
 in the optical spectrum of a galaxy, from the accumulation of absorption
 lines of mainly ionised metals. As the opacity increases with decreasing
 stellar temperature, the value of D$_n$4000 is largest for old and metal-rich stellar
 populations. Stellar population synthesis models indicate that D$_n$4000 is
 mostly sensitive to age, except for sub-solar metallicity populations older
 than 1 Gyr, where it is also strongly dependent on metallicity
 \citep{kauff03a}.

Interestingly, the variation of \hd with D$_n$4000 is not sensitive to
metallicity or dust extinction, thus providing a good measure of SFH,
 such that for a given D$_n$4000, stronger \hd
implies that the galaxy has experienced a recent burst in star
formation \citep{kauff03a}. In particular, the right panel of
\citeauthor{kauff03a}'s fig.~2 indicates that the combination of $\rm
H\delta\!>\!2\,\rm\,\AA$ and D$_n4000\!<\!1.5$ leads to stellar populations younger
than 1 to 3 Gyr for solar and 0.2 solar metallicity populations, respectively.

 \section{Velocity modulation of galaxy properties}
 \label{analysis}
 
 The key issue that we are trying to address in this paper is whether
 it is possible to distinguish the galaxies that have passed through
 the cluster core from those that are falling into it for the first
 time, based on observable parameters. DM simulations
 show that the backsplash population is most likely
 to be seen between $1-1.5$\rv and with velocities lower than the mean
 velocity of the cluster (e.g. \citealp{mamon04,gill05}). So, we
 stack 
 together all the galaxies found in clusters within $2$\rv and $|\Delta
 v_{\rm LOS}| < 3\,\sigma_v$,
into four bins of projected radius of width 0.5$R_{\rm v}$, and into three bins of absolute LOS velocity: $|\Delta v_{\rm LOS}|/\sigma_v$
 $<$1, 1--2 and 2--3 respectively. Below we describe the features of these distributions
 for various galaxy properties with respect to one
 another and the field distribution.

 \subsection{Stellar mass}
 \label{sec:mass}
 
 In Fig.~\ref{lgm} we plot the distribution of stellar masses for the
 12 classes (3 scaled velocity bins in each of the 4 scaled radial
 bins) of galaxies as described above.  Due to the flux limit of SDSS,
 the stellar mass distribution of galaxies in Fig.~\ref{lgm} appears to
 be centred around $M^{*}\!=\!3\!\times\!10^{10}\,{\rm M_\odot}$ in the
 local Universe \citep[also see][]{kauff03a}.  The most massive
 galaxies ($\log M^{*}/{\rm M_\odot}\!\geq\!11$) are mostly confined to the
 cores of the clusters. However, even in the core there seems to be a
 correlation between the relative LOS velocity of a galaxy with
 respect to the cluster and its stellar mass. 
 The Kolmogorov-Smirnov\footnote{The null hypothesis
   for the K-S test is that both the distributions being compared are
   drawn from the same parent distribution, \ie~if the two
   distributions are identical, the probability ($p$) of the null hypothesis
   being satisfied is unity.} (K-S henceforth) statistics show that
 the most massive
 galaxies in clusters have low velocities with respect to the cluster,
 and the high velocity galaxies show an affinity for the field
 distribution, which is skewed towards the low mass end. The detailed
 statistical comparison between different distributions is shown in
 Table~\ref{table:lgm} in Appendix.
 
 When the brightest cluster galaxies (BCGs) are excluded, there is no mass
 segregation in clusters relative to the field \citep{vonderlinden10}. To explore
 the inter-dependence of $M^*$, \rv and LOS velocity of galaxies in clusters,
 we repeated our analysis by excluding the BCGs from the clusters. Contrary to
 \citet{vonderlinden10}, we find that the different distributions still show
 similar trends: in the inner regions of clusters the most massive galaxies have
 lowest absolute LOS velocities (see Table~\ref{table:lgm-nobcg}). 

\begin{figure}
 \centering
\includegraphics[angle=-90,width=\hsize]{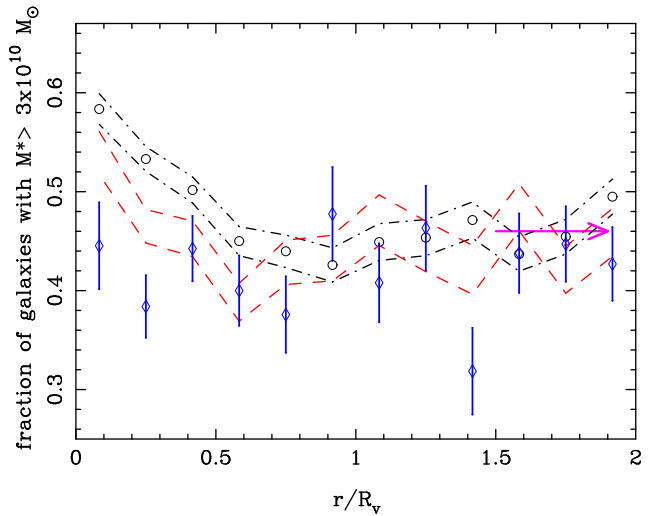}
\caption{The fraction of high stellar mass galaxies ($M^{*} \geq
  3\times10^{10}\,\rm {\rm M_\odot}$, with brightest cluster galaxies removed),
plotted as a
  function of scaled radius from the cluster centre. The low ($0-1\,\sigma_v$) and high
($>2\,\sigma_v$) absolute velocities (shown as \emph{black circles} and
\emph{blue diamonds} respectively), and the $\pm1\,\sigma$ range (from
binomial statistics) are shown as \emph{black dash-dotted lines}, \emph{red dashed lines}, and
\emph{blue error bars}, for the low, intermediate (1--2$\,\sigma_v$) and high
LOS absolute velocities, respectively. The \emph{magenta arrow} indicates the field value. }
 \label{frachimass}
 \end{figure}

 To analyse mass segregation in a different way, in Fig.~\ref{frachimass}
 we show the fraction of high-stellar mass galaxies as a function of projected
 radius in three bins of absolute LOS velocity, discarding BCGs.
Fig.~\ref{frachimass} confirms our impression that \emph{mass segregation is
  present, but only
at $R<0.5\,R_{\rm v}$, and only for the low- and intermediate $|v_{\rm LOS}|$ galaxies}.

The difference in low and high mass velocity distributions in
 the projected cluster core is not simply caused by the BCGs, which are excluded from
 Fig.~\ref{frachimass} (see also Table~\ref{table:lgm-nobcg}).
While \cite{biviano92} also found that high mass non-BCG galaxies had lower
absolute LOS velocities, we find that this is limited to galaxies within
$0.5\!R_{\rm v}$. In other words, we find that \emph{the high absolute LOS
velocity galaxies show no mass segregation}, in contrast with the global luminosity
segregation found by others (e.g. \citealp{ABM98,vonderlinden10}), when no cuts are made on
absolute LOS velocity (which we also find, since the high absolute LOS velocity galaxies
are much rarer than their intermediate and low velocity counterparts).
We discuss mass segregation amongst cluster galaxies further in \S\ref{masseg}.

 \subsection{Galaxy colours}
 \label{colour}
 \begin{figure*}
 \centering{
 {\rotatebox{270}{\epsfig{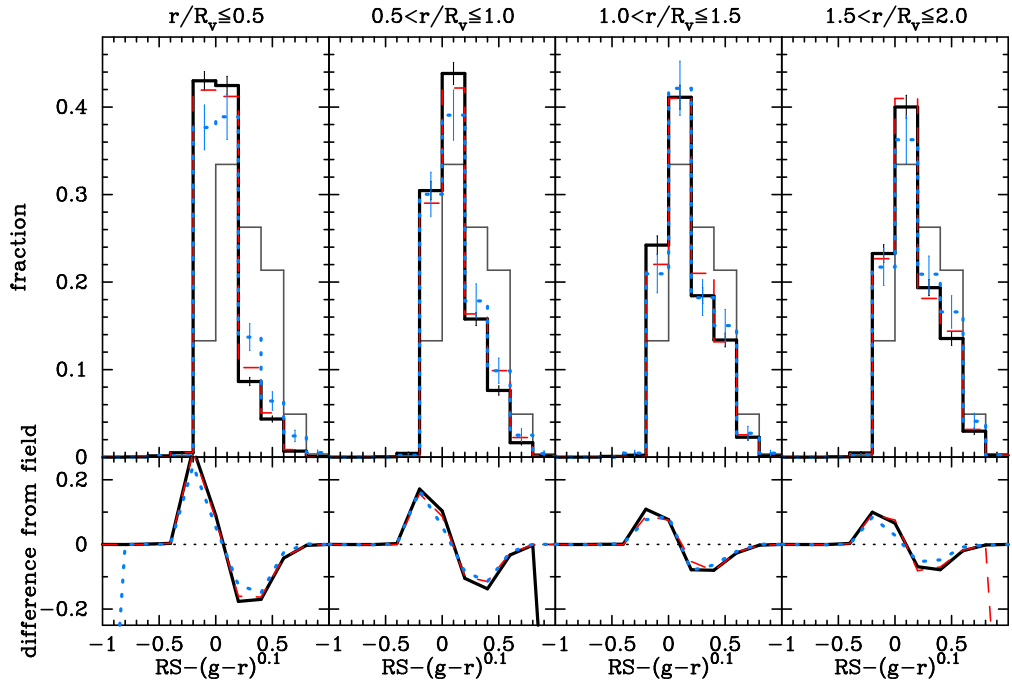}}}}
\caption{Same as Fig.~\ref{lgm} but for the ``colour offset'' of 
  cluster galaxies, i.e., the difference between the $(g\!-\!r)^{0.1}$
  colour of a galaxy and the Red Sequence colour for a galaxy of the
  same magnitude. Galaxies bluer than the fitted red sequence
 (see text) have positive
  values of the offset on the x-axis. The build-up of the blue tail of
  the distribution outwards from the cluster core is clearly
  seen. Interestingly, even in the outermost radial bin (1.5--2$R_{\rm v}$),
  there are more red sequence galaxies than the field. This shows
  that physical processes, dependent on environment, efficiently
  modify properties of galaxies in regions with densities in between
  the two extremes characterising cluster cores and the field. }
 \label{gr}
 \end{figure*}
  
 \begin{figure*}
 \centering{
 {\rotatebox{270}{\epsfig{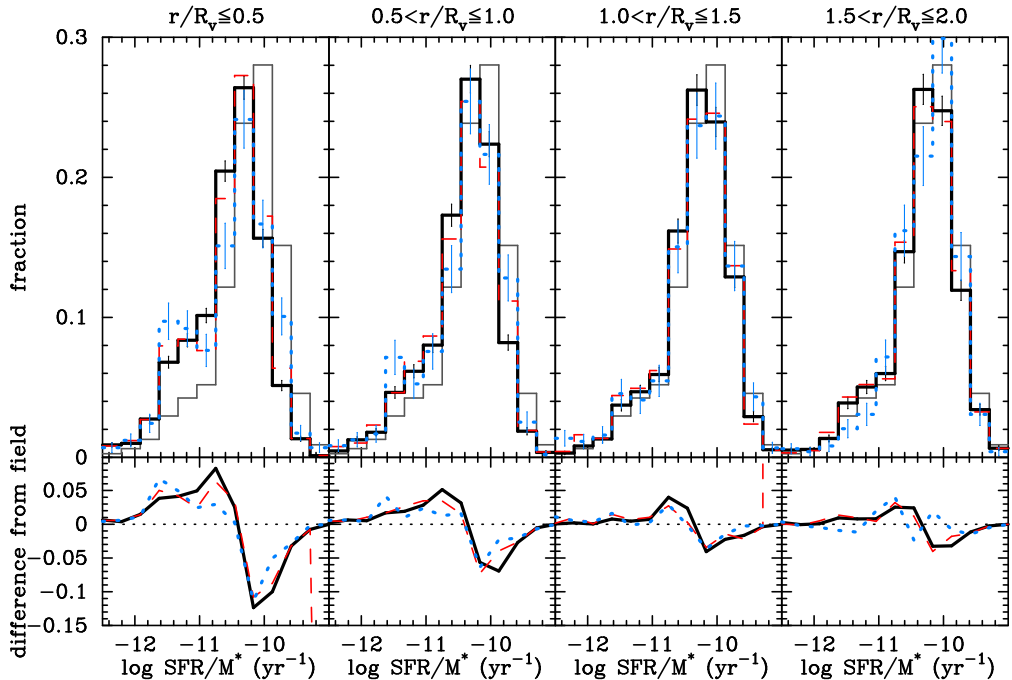}}}}
\caption{Same as Fig.~\ref{lgm} but for SFR/$M^*$.
  As expected, the cluster distributions are skewed towards
  the low \ssf\ end in the core region, but only a small tail of
  galaxies with low \ssf\ is seen in the outer radial bins. }
 \label{ssfr}
 \end{figure*}

 \begin{figure*}
 \centering
 \rotatebox{270}{\epsfig{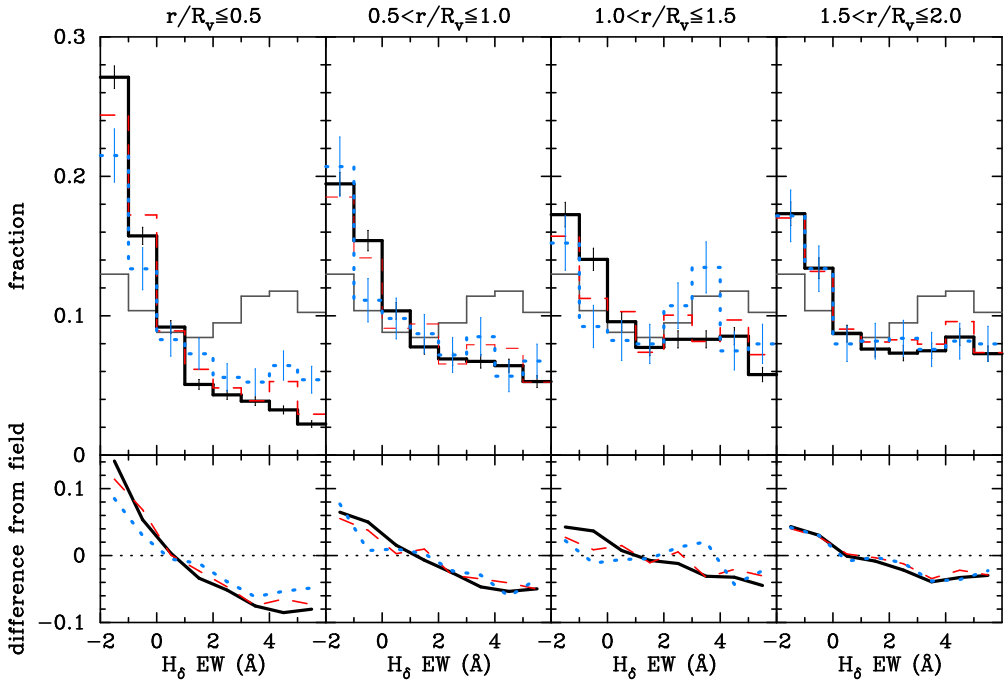}}
 \caption{Same as Fig.~\ref{lgm} but for \ehd. This plot
     complements the results in Fig.~\ref{hd}. The field distribution
   shows a clear bimodality around $2$\,\AA. The galaxies in the core of
   the cluster, most of which are evolving passively, do not show
   absorption in H$\delta$. It is interesting to note that the high
   velocity galaxies in the innermost bins, which are likely to be the
   newest entrants in the core, are also the ones which are most
   likely to have experienced a burst of star formation as indicated
   by relatively high \ehd. This trend in velocity classes continues
   out to 2$R_{\rm v}$, where the difference between the field and the low
   velocity galaxies is minimum. The difference between the
   highest ({\it{blue}}) and the lowest velocity ({\it{black}})
   galaxies is statistically significant in all radial bins out to
   1.5\,$R_{\rm v}$.}
 \label{hd}
 \end{figure*}

 \begin{figure*}
   \centering{ {\rotatebox{270}{\epsfig{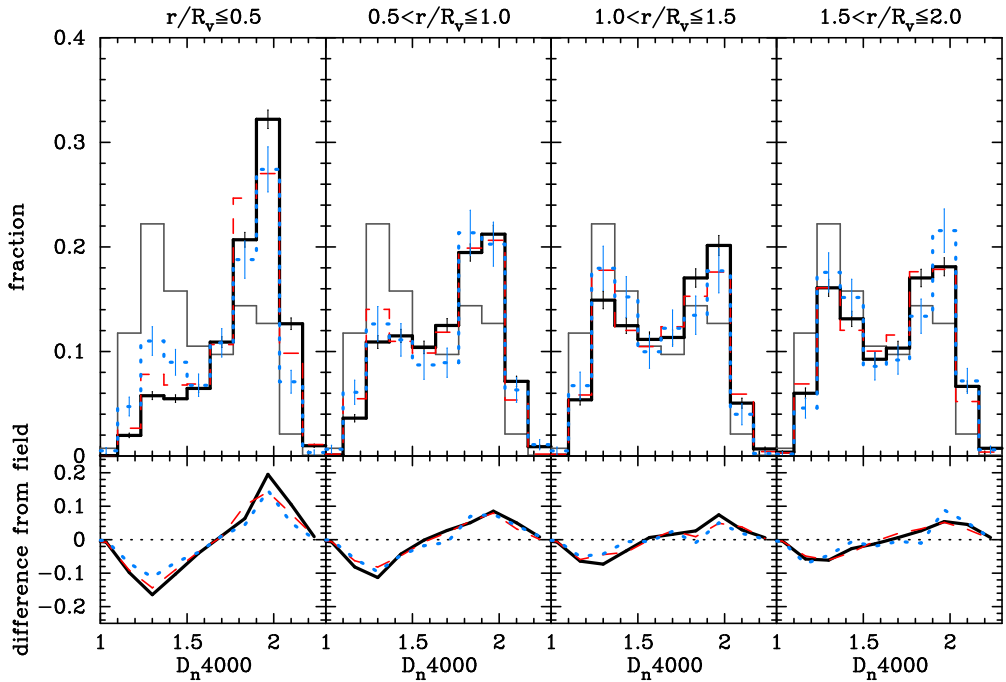}}}}
   \caption{Same as Fig.~\ref{lgm} but for D$_n$4000. The field distribution
     is bimodal with a minimum at D$_n$4000\break$\sim\!1.5$. The
     passively evolving galaxies in the cluster core are old and
     metal-rich, hence their distribution peaks at the high end of the
     D$_n$4000. However, the high velocity galaxies in the cluster core
     (dotted blue line) have a distribution similar to that of field galaxies. 
     The trend continues
     out to 1.5\,$R_{\rm v}$, beyond which the distributions almost merge with
     the field. The K-S statistics show the distributions of the high 
     and low velocity galaxies to be statistically different at all radial 
     distances within 1.5$R_{\rm v}$. }
 \label{d4}
 \end{figure*}

  \begin{figure*}
  \centering{
  {\rotatebox{270}{\epsfig{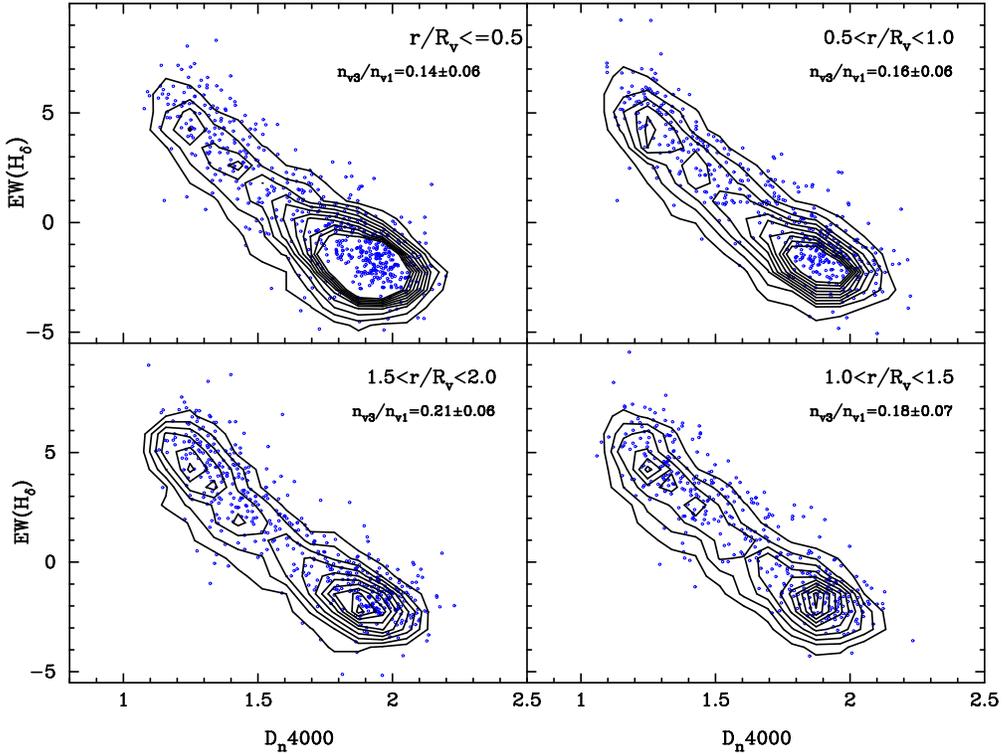}}}}
\caption{The distribution of galaxies having the highest
  ($|v_{\rm LOS}|/\sigma_{v} = 2-3$; \emph{blue points}) and the lowest
  ($|v_{\rm LOS}|<\sigma_{v}$; \emph{black contours}) absolute LOS velocity relative to the cluster,
  in bins of increasing (scaled) radius (\emph{clockwise} from top
  left).  The innermost contour represents a density of 100,
  decreasing by 10 at every subsequent level.  The fractions in the
  top right corner of each panel show the ratio of the numbers in the
  two classes of galaxies shown, and the scatter in it. Barring the
  innermost radial bin, the distribution of high velocity galaxies
  remains the same, although the distribution of the virialised galaxies tends
  to become bimodal with increasing (scaled) radius. }
  \label{grid}
  \end{figure*}

 Broadband photometric colours are widely used as proxies for the star
 formation activity of galaxies. In this paper, we use the `colour
 offset', which is the difference between the $(g\!-\!r)^{0.1}$
 (extinction-corrected) colour of a galaxy and that of the red sequence for a galaxy of the
 same magnitude, where the red sequence is fitted by
 stacking all the cluster galaxies together.
  In Fig.~\ref{gr} we plot the distribution of the colour offset for all 12 classes
 of galaxies.

 The low velocity galaxies in cluster cores are found to have the
 reddest colours, indicative of their passive evolution. But
 interestingly, even within the cores, the difference between the
 colours of high and low velocity galaxies indicates statistically 
 significant different distributions (see Table~\ref{table:gr}). 
 Out to 2$R_{\rm v}$, the distribution of the colour
 offset for the field galaxies is very different from that of the
 galaxies residing in and around clusters. This might suggest that the
 broadband colours could be very easily modulated by slight changes in
 the local and/or global environment, and are hence not very well
 suited for the study of evolutionary trends amongst galaxies
(see also \citealp*{mahajan10,mahajan11}).
As discussed below, on the
 outskirts of clusters (1--1.5$R_{\rm v}$) the trends seen for broadband colours are
 consistent with those seen for spectroscopic parameters.
   
\subsection{Diagnostics of specific star formation rate and star formation history}

 It has been shown  that the SFR of galaxies is
 a function of their stellar mass \citep[\eg][]{noeske07}. So in this
 work we prefer to use the specific star formation rate (SFR/$M^*$) to quantify
 the {\emph{current}} star formation activity in a galaxy. 
 In Fig.~\ref{ssfr}, we plot the distributions of SFR/$M^*$, as estimated by
 \citet{b04}, for all the galaxies in the spectroscopic galaxy catalogue of the SDSS
 DR4.

 As expected, the cluster galaxies within 0.5$\,$\rv have their
 \ssf~distributions heavily skewed towards the lower end, indicating passive evolution.
 Interestingly though, the high end of the values of SFR/$M^*$ in the cluster core, and
 almost at all radii out to $1.5\,R_{\rm v}$, mostly belongs to galaxies with
 high velocities. K-S statistics show
 the difference between the lowest and the highest velocity galaxy
 distributions to be statistically
 significant in the core ($<\!0.5\,R_{\rm v}$) and on the cluster outskirts
 ($1\!-\!1.5\,R_{\rm v}$; see Table~\ref{table:ssfr}). The field distribution is
 significantly different from all other classes of galaxies, becoming
 indistinguishable from the high velocity galaxies only in the outermost
 radial bin.

Figs.~\ref{hd} and \ref{d4} show the distributions of \ehd\footnote{Note that the
 negative values of \ehd~do not imply \hd line {\emph{emission}} in the passive galaxies,
 because the EW(H$\delta$) values are measured from the pseudo-continuum
 bands, and so do not always represent the `true' flux in the \hd
 line. The pseudo-continuum bands are required because in the spectra
 of passive galaxies, the position of several metal lines may coincide
 with the \hd line, making it virtually impossible to measure the
 `true' flux (G. Kauffmann, priv. comm.; see also \citealp{worthey97} and \citealp{kauff03a}).} and  
 D$_n$4000 for the galaxies in 12 different cells of projected phase space 
(three scaled absolute LOS velocities times four scaled projected radii), 
and compare them with the distribution of field galaxies.
The \ehd~distribution of the cluster galaxies is strikingly
 different from that of the field galaxies, the latter showing a bimodality at
 \ehd\,$\sim\!2$\,\,\AA~(also see Table~\ref{table:hd}).
 In the cluster cores, the galaxies with highest velocities, which are
 most likely to have fallen in recently (Table~1), are the ones that also show
 \hd in absorption (\ehd\,$>\!0$). This suggests that their likelihood of
 experiencing a burst of star formation within the last $\sim\!1-3$\,Gyr is
 relatively higher than their low velocity counterparts. The K-S test
 shows that the difference between the \ehd~distributions of the highest
 and the lowest absolute LOS velocity galaxies is statistically significant at
 all radii within $1.5\,R_{\rm v}$ (Table~\ref{table:hd}).
 
The amplitudes of the 4000$\,\rm \AA$ break in SDSS galaxy spectra
 show very different distributions
 for the highest and lowest classes of $|v_{\rm LOS}|$, both in the core ($<\!0.5\,R_{\rm v}$) and on
 the periphery ($1\!-\!1.5\,R_{\rm v}$) of the clusters. However, the
 distributions of  D$_n$4000 in the $0.5\!-\!1\,R_{\rm v}$ bin are only very marginally
 dissimilar, with $\sim\!10$\% probability of the null hypothesis being
 satisfied (see Table~\ref{table:d4}). 

 We now take advantage of the fact that together H$\delta$
 and D$_n$4000 yield a good constraint on the 
 SFHs \citep{kauff03a} of galaxies. In Fig.~\ref{grid} we show 
 the SFHs of galaxies with different absolute LOS velocities in bins of
 increasing cluster-centric radius. For clarity, we only plot the two
 extreme absolute LOS velocity classes ($|v_{\rm LOS}|<\sigma_{v}$ and
 $|v_{\rm LOS}|/\sigma_{v} = 2-3$). The distribution of the low $|v_{\rm LOS}|$ 
 galaxies, in this plot of \ehd~vs D$_n$4000, is dominated by the passive sequence (lower
 right) at low projected radii, and becomes increasingly
 bimodal with increasing cluster-centric radius. The same evolution occurs
 for the high $|v_{\rm LOS}|$ galaxies, but is less bimodal at large
 projected radius.
 On the other hand, the high absolute LOS velocity galaxies span the entire
 available parameter space equally at all distances. 
According to the stellar population study of \citeauthor{kauff03a}
 (see their fig.~3), this is an indication that the SFH is more continuous in the low
$|v_{\rm LOS}|$ galaxies and star formation occurs more in bursts in the high velocity galaxies.
 We discuss these trends in the context of the relative fraction of the infalling and
 backsplash galaxies, and the fraction of star-forming galaxies as a
 function of cluster-centric distance in \S\ref{discussion}.

 We now refer to galaxies with EW(H$\delta$)\,$>$\,2 and
 D$_n$4000$\,<\!1.5$ as \emph{Galaxies with Ongoing or Recent 
 Efficient Star formation} (GORES), following the 
 assumption that a galaxy that experienced efficient star formation
$\lesssim\!1-3$\,Gyr ago is likely to show these features prominently in its
 spectrum. Our choice of EW(H$\delta$) and D$_n$4000 thresholds are based
 on the bimodality seen in the field distributions (Figs.~\ref{hd} and \ref{d4}).
   
\begin{figure}
\centering
\includegraphics[width=\hsize]{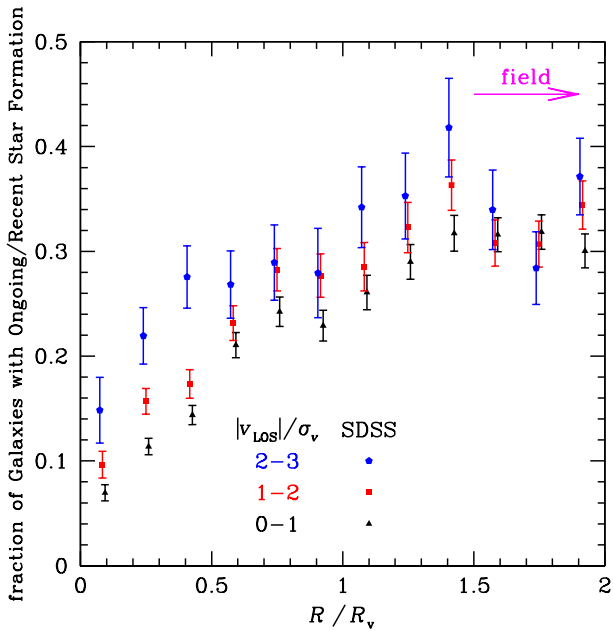}
\caption{The fraction of Galaxies with Ongoing or Recent Efficient Star
  Formation (GORES, EW(H$\delta$)\,$>$\,$2\,\AA$ and
 D$_n$4000$\,<\!1.5$) versus projected radius in bins of absolute LOS
 velocity: 
$0-1\,\sigma_v$ (\emph{black triangles}),
$1-2\,\sigma_v$ (\emph{red squares}), and
$> 2\,\sigma_v$ (\emph{blue pentagons}), slightly shifted along the $x$-axis
 for clarity.
The \emph{magenta arrow} indicates the GORES fraction in the field
($R>3\,R_{\rm v}$).
\label{fGORESvsR}}
\end{figure}

 Fig.~\ref{fGORESvsR} shows how the fraction of GORES, $f_{\rm GORES}$,
 varies with projected 
 radius and absolute LOS velocity.
The generally increasing radial trend of 
$f_{\rm GORES}$ is clearly modulated by $|v_{\rm LOS}|$: at
all projected radii, the value of $f_{\rm GORES}$ is higher (lower) for
high (low)  $|v_{\rm LOS}|$.

\section{Modelling the velocity modulation}
\label{models}

In this section, we present simple models to quantify the observed fraction
 of GORES by combining theoretical predictions with
 the suite of galaxy properties analysed in \S\ref{analysis}.

We build two simple models of the frequency of GORES, one in which their
frequency only depends on the \emph{class} of galaxies, i.e. whether they are part of the
virialised, infall or backsplash population, and second, where the frequency
 of GORES depends upon class \emph{and} the distance from the centre of the cluster.
We define the three classes of galaxies in radial phase space (i.e. $[r,v_r]$, where $r$
is the 3D radius and $v_r$ the radial velocity) using the $z\!=\!0$ output of
a cosmological simulation. We also use this simulation to project these
quantities onto the
observable \emph{projected phase space}, $[R, |v_{\rm LOS}|]$, where $R$ is the
projected radius.

\subsection{Cosmological simulation}
 \label{simulations}
 
 We use the high-resolution 
 cosmological hydrodynamical simulation run by \cite{borgani04}. To summarise, the
 simulation assumes a cosmological model with $\Omega_0\!=\!0.3$,
 $\Omega_\Lambda\!=\!0.7$, $\Omega_{\rm b}\!=\!0.039$, $h\!=\!0.7$,
 and $\sigma_8\!=\!0.8$. The box size is $L\!=\!192\,h^{-1}$ Mpc. The
 simulation uses $480^3$ DM particles and (initially) as many
 gas particles, for a DM particle mass of
 $4.62\!\times\!10^9\,h^{-1} {\rm M_\odot}$.  The softening length was set
 to $22.5 \, h^{-1} \, \rm comoving\ kpc$ until redshift $z\!=\!2$ and fixed
 for subsequent epochs (i.e., at $7.5\, h^{-1}$ kpc). DM halos were
 identified by \citeauthor{borgani04} at redshift $z\!=\!0$ by applying a
 standard FoF analysis to the DM
 particle set, with linking length 0.15 times the mean inter-particle
 distance. After the FoF identification, the centre of the halo was
 set to the position of its most bound particle. A spherical
 overdensity criterion was then applied to determine the virial
 radius, $R_{200}$ of each halo.
 In this manner, 117 halos were identified within the simulated
 volume, of which 105 have virial mass $M_{200}$ greater than
 $10^{14}\,h^{-1} {\rm M_\odot}$. To save computing time, the subsequent
 analysis was performed on a random subsample of roughly 2 million
 particles selected from the whole sample of $480^3$ particles.  
 We deliberately chose to analyse DM particles rather than galaxies in
 the simulations, as we do not trust the galaxy distribution, with a
 homogeneous core found in these and other hydrodynamic cosmological
 simulations, as they conflict with the observations
 \citep{Saro+06}. Moreover, the particle data provide much
 better statistics.

A sample of 93 non-bimodal mock clusters were extracted from these
simulations by \citet*{mbm10}, who placed their observer at $90 \, h^{-1} \,
\rm Mpc$ of each mock cluster and added the peculiar motions to the distance
with the Hubble constant value  ($H_0 = 100 \,\rm km \,s^{-1} \, Mpc^{-1}$)
corresponding to the units of positions ($h^{-1} \, \rm kpc$). This yields fairly regular
clusters that should resemble those of \cite{yang07} (in the latter case, their high
threshold causes multimodal clusters to be fragmented into several).
We repeated the same extraction, this time with by placing the observer at
 an infinite distance of each cluster (cylindrical projection).

\subsection{Classification schemes}
\label{classif}

Ideally, one should define the virial, infall and backsplash classes by
following the orbits of the particles \citep{gill05}, but this task is
beyond the scope of the present article. Instead, we define the classes using
the present-day radial phase space distribution of particles.
 Particles that are outside the virial radius, and travelling at velocities
 lower than the critical velocity, are falling into the cluster for
 the first time, while particles outside the virial radius with higher
 velocities must have crossed the cluster at least once.

\cite*{SLM04} have analysed the distribution of DM particles in the
radial phase space $(r,v_r)$ at $z\!=\!0$, for a cluster-mass halo in a
dissipationless cosmological $n$-body simulation ($256^3$ particles in
a box of $150 \, h^{-1} \, \rm Mpc$ width). Their fig.~1 (see also
\citealp{mamon04}) suggests that the separation of particles belonging to the infall,
backsplash and virialised populations can be written with a critical
 radial velocity delimiting the infall regime from the other two classes:
 \begin{equation}
{  v_{r,{\rm crit}}\over V_{\rm v}}= -1.8+ 1.06\,\left( {r\over R_{\rm v}}\right) \ .
\label{vrcrit}
 \end{equation}

\begin{figure}
\psfig{file=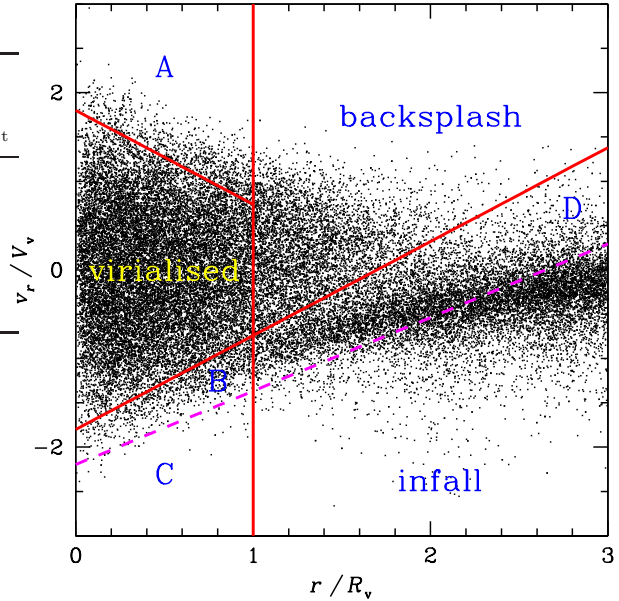,width=\hsize}
\caption{Radial phase space distribution of dark matter particles
  of the stack of 93 regular mock clusters from the cosmological hydrodynamical
  simulation of \citet{borgani04}. The units of radius and radial velocity
  are the virial radius $r_{100}$ and the circular velocity at that radius,
 respectively. The critical velocity separating
  infall from backsplash population ({\it{long diagonal red line}}) is
given in Eq.~\ref{vrcrit}. The {\it{short diagonal line}} represents the
  negative critical velocity threshold (see \S\ref{models}). For
  clarity, only 1 out of 550 particles of the original simulation is plotted.
Region~A is virialised in Schemes~0 and 1, infall in Scheme~3, and otherwise
backsplash. Region~B is virialised in Scheme~0, infall in Schemes 1 to 3, otherwise
backsplash. Region~C is virialised in Scheme~0, backsplash in Scheme~4, otherwise infall.
Region~D is backsplash in Scheme~5, otherwise infall (see Table~\ref{schemes}).}
\label{rvr}
\end{figure}

Fig.~\ref{rvr} shows the distribution of DM particles in the stack of 93 clusters 
in radial phase space
$(r,v_r)$.
Overplotted are the cuts of \cite{SLM04} designed to separate the
virialised, infall, and 
backsplash classes of particles (Eq.~\ref{vrcrit}). 
The particles within the virial radius with large absolute radial velocities have
uncertain classes. 

We devise several \emph{Schemes} to identify the three
classes of particles (we refer to fig.~6 in \citealp{Bertschinger85} for a
schematic view of the trajectories of particles in radial phase space).
 Table~\ref{schemes} summarises these schemes.
The rapidly infalling particles with $r < r_{\rm v}$ might represent the
low-end tail of the radial velocity distribution of the virial class, which
we denote by Scheme~0, or might indeed be part of
the infall class (Schemes 1 to 3),
or may even be mainly populated by
particles on their second infall, i.e. backsplash class (Scheme~4). The
rapidly expanding particles with $r < 
r_{\rm v}$ might again represent the high-end tail of the radial velocity
distribution of the virial class (Schemes~0 and 1), or the backsplash class
(Schemes 2 and 4), or the infall class (if the star forming properties of galaxies
in this region of radial phase space reflect those of the rapidly infalling
class within the virial radius, Scheme~3).
Scheme~0 corresponds to maximum virialisation while Scheme~3 corresponds to
maximum infall.
 
Although Fig.~\ref{rvr} is based upon a very different simulation (hydrodynamical
instead of dissipationless and with three times better mass resolution) and on a
stacked halo instead of a single one, it shows that the critical radial velocity of
Eq.~\ref{vrcrit} is fully adequate to distinguish between infalling and
backsplash particles in the stacked mock cluster obtained from the
cosmological hydrodynamical simulation analysed here.

One may argue that the red solid line separating the infall and backsplash classes is
too high and that we may be missing an important fraction of backsplash
particles at high $r$ and low $v_r$. We thus created a \emph{maximum backsplash}
 Scheme~5, where we consider the critical velocity
\begin{equation}
{v_{r,{\rm crit}}^{(2)} \over V_{\rm v}} = -2.2+ 0.83\,\left ({r \over R_{\rm v}}\right).
\label{vrcrit2}
\end{equation}
This is shown as the magenta dashed line in Fig.~\ref{rvr}: particles with $v_r
< v_{r,\rm crit}^{(2)}$ are infalling, other particles are backsplash or virial
according to Scheme~4 applied with $v_{r,\rm crit}$.

\begin{table}
\begin{center}
\caption{Schemes of particle classes from positions in radial phase
  space\label{schemes}}
\tabcolsep 3pt
\begin{tabular}{lccccc}
\hline
Scheme & Region & A & B & C & D \\
& $r/R_{\rm v}$ & $<1$ & $<1$ & $<1$ & $>1$ \\
& $v_r$ & $>- v_{r,\rm crit}$ & $v_{r,\rm crit}^{(2)} \leftrightarrow  v_{r,\rm crit}$ &
$<v_{r,\rm crit}^{(2)}$ &  $v_{r,\rm crit}^{(2)} \leftrightarrow  v_{r,\rm crit}$ \\
\hline
0 & & virial & virial & virial & infall \\
1 & & virial & infall & infall & infall \\
2 & & backsplash & infall & infall & infall \\
3 & & infall & infall & infall & infall \\
4 & & backsplash & backsplash & backsplash & infall \\
5 & & backsplash & backsplash & infall & backsplash \\
\hline 
\ref{schemes}
\end{tabular} 
\end{center}

Note: The letters A to D correspond to the regions in Fig.~\ref{rvr}. The
symbol $\leftrightarrow$ means `in between'.
\end{table}

The cosmological simulation provides us with a unique way of measuring
the frequency of the three classes of particles (galaxies) in
projected phase space.  We translate velocities from $V_{\rm v}=V_{100}$ to the
observable cluster velocity dispersion, $\sigma_v$, noting that the
stacked cluster has a velocity dispersion (limited to the aperture
$r\!<\!R_{\rm v}$) of
 $0.65\,V_{\rm v}$, as measured in the hydrodynamical simulation by
\cite{mbm10}, and within 5\% of what is expected for isotropic 
NFW models of clusters (Appendix~A of \citealp{MM07}).

\begin{figure}
\psfig{file=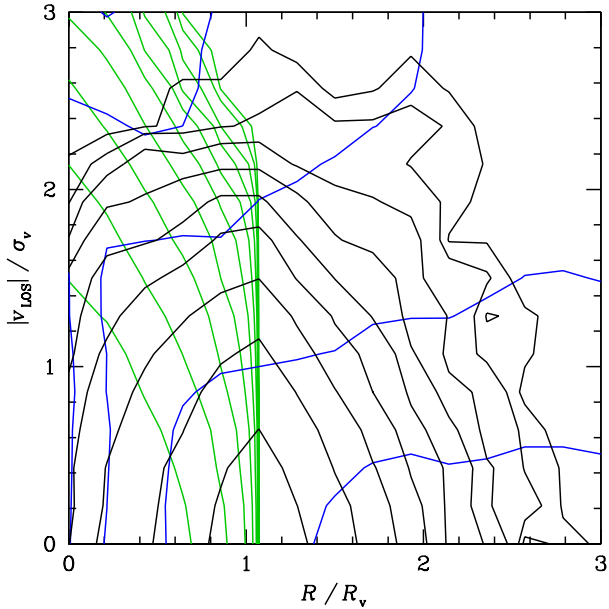,width=\hsize}
\caption{Density in projected phase space
of virial (\emph{green}), infall (\emph{blue}) and backsplash
 (\emph{black contours}) particles of the stacked mock cluster
 with our preferred model (Scheme 3, see Fig.~\ref{rvr}).
 The \emph{contours} are logarithmically spaced by
 a factor of 2.3, and the same set of contours is used for all three classes:
the highest contours are on the lower left (virial), lower right (infall) and
bottom (backsplash), where the latter two classes only reach the 2nd highest
contour of the virial class.
}
\label{Rvconts}
\end{figure}
Fig.~\ref{Rvconts} shows the fractions of the three classes in projected
phase space.
The virialised particles are necessarily at radii $r\!<\!R_{\rm v}$, the infalling
particles prefer large projected distances and high velocities, while the 
 backsplash particles are mostly found just around the virial radius
 and have \emph{low absolute LOS velocities}. This preference of backsplash
 galaxies  for low
 $|v_{\rm LOS}|$ is an immediate consequence of their lower radial velocities
 (\S\ref{intro}) and has also been noted by \cite{gill05}.

\begin{figure}
 \psfig{file=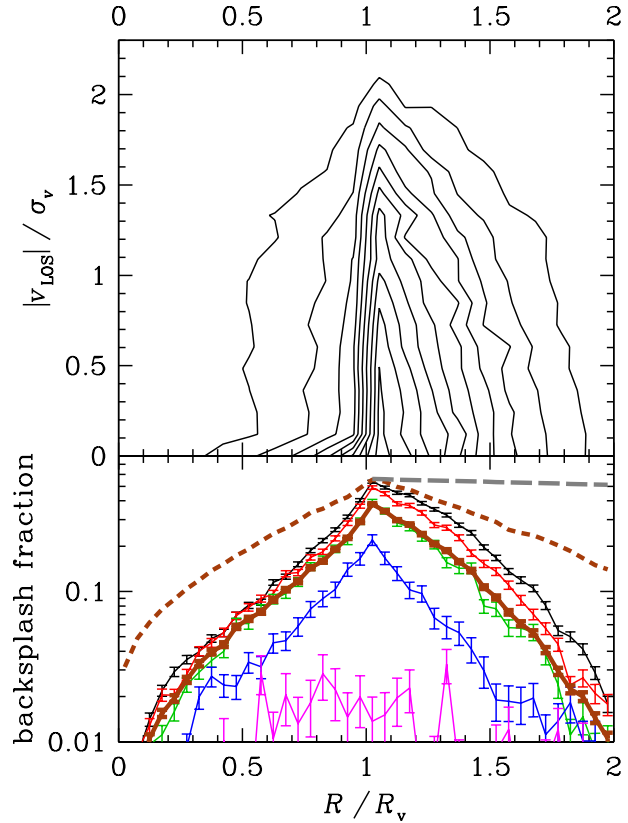,width=\hsize}
 \caption{\emph{Upper panel}: contours in projected phase space of the 
fraction of particles  of the stacked mock cluster that contribute to the backsplash class 
(Scheme 3; see Fig.~\ref{rvr}). \emph{Contours} are linearly spaced from 0.05 to 0.5
 going downwards. \emph{Lower panel}: fraction of backsplash particles in bins of
 absolute LOS velocity: 0--0.5 (\emph{top, black}), 0.5--1.0 (\emph{red}), 1.0--1.5
 (\emph{green}), 1.5--2.0 (\emph{blue}), and $>\!2.0\,\sigma_v$ (\emph{bottom,
 magenta}), and summed over all LOS velocities $< 3\,\sigma$ (\emph{thick brown}).
 The \emph{dashed thick brown curve} shows the maximum backsplash fraction
 (for $|v_{\rm LOS}|<3\,\sigma_v$)
 from Scheme~5 while the \emph{long-dashed thick grey line} is the relation
 deduced from observations by \citet{Pimbblet11} (see \S\ref{bspid}). }
 \label{bspfrac}
 \end{figure}
In the top panel of  
Fig.~\ref{bspfrac}, we show the contours for the fraction of particles that 
 are backsplash, while the bottom panel provides the fraction of backsplash
 particles versus projected radius in wide bins of absolute LOS velocity.
The two plots of Fig.~\ref{bspfrac} indicate that the fraction of backsplash
particles reaches a maximum of 54\% just outside the virial radius at very
low $|v_{\rm LOS}|$ for Scheme~3 (75\% for Scheme~5). At high $|v_{\rm LOS}|$, 
the fraction of backsplash particles is symmetric relative to the virial
radius. However, at low $|v_{\rm LOS}|$, the fraction is skewed towards higher
values beyond the virial radius (see, e.g. curve representing all particles
 with $|v_{\rm LOS}| < 3\,\sigma_v$ in the bottom panel of Fig.~\ref{bspfrac}).
Indeed, at projected radii smaller than the virial radius, the backsplash
particles must be in the foreground or background of the virial sphere, and
geometric effects typically cause their 3D velocities to be
aligned with the line of sight, making it difficult to obtain small
absolute LOS velocities.
These trends are confirmed in Table~\ref{tabfracs}, which lists the fractions of
the three classes of particles in cells of projected phase space.
\begin{table*}
\begin{center}
\caption{The fraction of virialised (v) infall (i) and backsplash (b) particles
  of the stacked mock cluster in  
cells of projected phase space with Scheme~3}
\begin{tabular}{cccccccccccccccc}
\hline
$r/R_{\rm v}$ & \multicolumn{3}{c}{0--0.5}
& & \multicolumn{3}{c}{0.5--1} & & \multicolumn{3}{c}{1--1.5} & &
\multicolumn{3}{c}{1.5--2} \\
\cline{2-4}
\cline{6-8}
\cline{10-12}
\cline{14-16}
$|v_{\rm LOS}|/\sigma_v$ & v & i & b & & v & i & b & & v & i & b & & v & i & b \\
\hline
0--1 & 0.89 & 0.08 & 0.03 & & 0.49 & 0.33 & 0.18 & &    0.00 & 0.65 & 0.35  & &    0.00 & 0.93 & 0.07 \\ 
1--2 & 0.83 & 0.16 & 0.02 & & 0.40 & 0.48 & 0.11 & &    0.00 & 0.81 & 0.19  & &    0.00 & 0.97 & 0.03 \\ 
2--3 & 0.69 & 0.31 & 0.00 & & 0.16 & 0.83 & 0.01 & &    0.00 & 0.99 & 0.01  & &    0.00 & 1.00 & 0.00 \\ 
\hline
\label{tabfracs}
\end{tabular}
\end{center}

\end{table*}

\subsection{Models}
\label{model12}

\subsubsection{Global deprojection}
\label{deprojsec}

We first directly deproject the observed fractions of GORES, regardless of
the LOS velocity.
Expressing the global (summing over the velocity bins) observed fraction of GORES as
$g_{\rm GORES}(R) = N_{\rm GORES}(R)/N_{\rm tot}(R)$, we can deduce the
surface densities, $\Sigma(R)=N(R)/(2\,\pi\, R\, dR)$ of GORES and of all galaxies.
\begin{figure}
\centering
\includegraphics[width=\hsize]{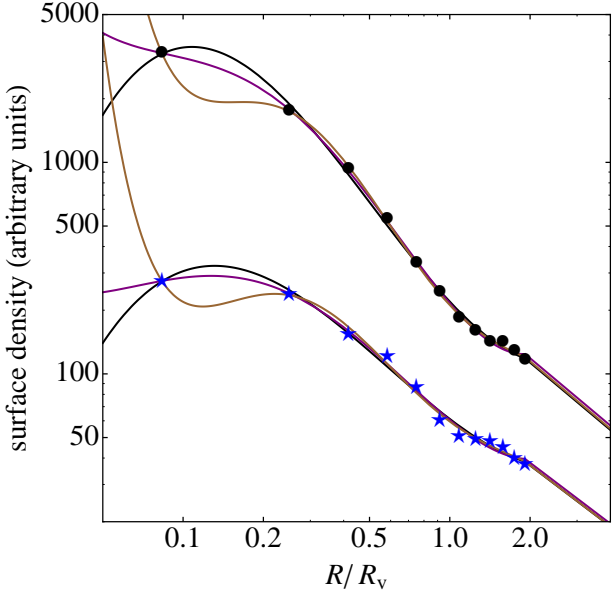}
\caption{Surface density of all galaxies (\emph{top}) and of GORES
 (\emph{bottom}). The \emph{symbols} are the data (\emph{filled black circles}
 for all galaxies and \emph{blue stars} for GORES). The \emph{black},
 \emph{purple}, and \emph{brown} curves represent the polynomial fits of
 $\log \Sigma$ vs $\log R$ for orders 3, 4, and 5, respectively. The data were
 extrapolated beyond $2\,R_{\rm v}$ with power-law fits to the outermost three
 data points.} 
\label{sdens}
\end{figure}
Since the projected number profiles $N_{\rm GORES}(R)$ and $N_{\rm tot}(R)$
 are fairly noisy, we fit a polynomial to $\log \Sigma$ vs. $\log R$ and
 extrapolate the data beyond the last point by a power-law, whose slope is fit
 on the last three points. Fig.~\ref{sdens} shows the surface density profiles
 and their polynomial fits. While the fits diverge dramatically at low $R$, they
 match remarkably well at high $R$ (better than analogous fits of
 $\log \Sigma$ vs. $R$, not shown in Fig.~\ref{sdens} for clarity).
One can now perform Abel inversion to deduce the space densities 
\begin{equation}
\nu(r)=-{1\over \pi} \int_r^\infty ({\rm d} \Sigma/{\rm d}
R)/\sqrt{R^2-r^2}\,{\rm d} R
\label{Abel}
\end{equation}
of GORES and all galaxies, and finally obtain the
deprojected fraction of GORES by dividing the two space densities.
Fig.~\ref{deproj} shows the resultant deprojected fraction of GORES,
obtained with the orders 3 to 5 polynomial fits of $\log \Sigma$ vs.
 $\log R$ (Fig.~\ref{sdens}).
Fig.~\ref{deproj} also shows a $\chi^2$ fit of
Eq.~\ref{modelfit} to the deprojected fraction of GORES obtained
from the order 4 polynomial fit of  $\log \Sigma$ vs. $\log R$, using
\begin{figure}
\centering
\includegraphics[width=\hsize]{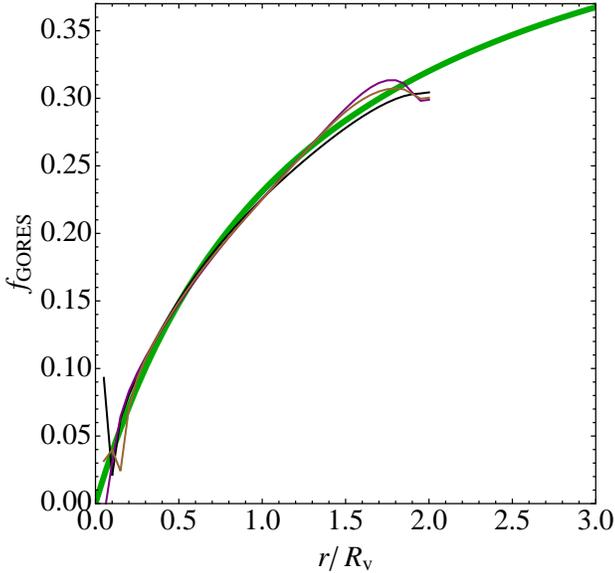}
\caption{
 Deprojected fractions of GORES (using Eq.~\ref{Abel}). \emph{Thin curves}
 show the results obtained by polynomial fits of the log surface density vs
 $\log R$ (orders 3, 4, and 5 in \emph{black}, \emph{purple} and \emph{brown},
 respectively). The \emph{thick green curve} represents Eq.~\ref{modelfit} with
 $f_0=0.52$ and $a = 1.26\,R_{\rm v}$, obtained by a $\chi^2$ fit to the order
 4 polynomial fit of $\log \Sigma$ vs $\log R$ (\emph{purple curve}), with $r$
 linearly spaced between 0.05 and $2\,R_{\rm v}$.
\label{deproj}}
\end{figure}
\begin{equation}
f_{\rm GORES} (r) = f_0 \,{r\over r + a} \ .
\label{modelfit}
\end{equation}
We see that the simple, saturated linear model provided in
Eq.~\ref{modelfit}, with asymptotic GORES fraction $f_0 = 0.52$ and
\emph{quenching radius for efficient star formation} (hereafter quenching radius)
$a=1.26\,R_{\rm v}$, is a decent
representation of the deprojected profile for $f_{\rm GORES}$.
While this deprojection leads to an asymptotic value of $f_{\rm GORES} =
0.52$, Eq.~\ref{modelfit} yields $f_{\rm GORES}=0.44$ at our typical
minimum projected radius for field galaxies ($6.7\,R_{\rm v}$), very close to
their observed fraction of 0.45.

Note that although the constraints from potential SDSS fibre collisions leads to
underestimated surface density profiles at low projected radii, this bias in
surface density should affect both the total and GORES surface densities in a
similar (multiplicative) way (Eq.~\ref{Abel}). Hence the effect of fibre collisions on 
the ratio of GORES to total space densities should be minor.

We now refine the models accounting for the LOS velocity modulation of
$f_{\rm GORES}(R)$, using the dynamical classes of the
galaxies (virial, infall, and backsplash).

 \subsubsection{Model 1: Constant recent starburst fraction per class}
\label{model1}

 Suppose that for each of the three classes of galaxies (virial, infall and
 backsplash), the fraction of galaxies which experienced a recent ($\lesssim\!1-3$ Gyr)
 starburst (GORES) is independent of the 3D radius and radial velocity. Let these
 three fractions be $f_{\rm v}$, $f_{\rm i}$ and $f_{\rm b}$ respectively (we
 drop the subscript `GORES' from $f$ for clarity).

 The fraction of GORES in a cell $(R,v_z)$ of projected phase space can then be
 written as
 \begin{equation}
 g_{\rm GORES}(R_i,v_j) = \sum_\alpha f_\alpha\,p(\alpha|R_i,v_j) \ ,
\label{gcst}
 \end{equation}
 where, $p(\alpha|R_i,v_j)$ is the fraction of particles of class $\alpha$ within the cell
 $(R_i,v_j)$ of projected phase space (as in Table~1, but with finer radial bins).

 One can then perform a $\chi^2$ fit between the observed and predicted
 fractions of GORES $g_{\rm GORES}(R_i,v_j)$ over all cells $(R_i,v_j)$, using the
 conditional probabilities for each class, $p(\alpha|R_i,v_j)$,
 derived from the cosmological
 simulation. We assume binomial error bars on the observed values of
 $g_{\rm GORES}(R_i,v_j)$ and that the $p(\alpha|R_i,v_j)$ derived from the cosmological
 simulation are perfectly known.
 Furthermore, we  use the six Schemes for the three populations, changing the
 class within the cluster virial sphere for the extreme velocities $|v_r| >
 |v_{r,\rm crit}(r)|$ as explained in the caption of Fig.~\ref{rvr}.
 Table~\ref{pops} shows the best fits for each of these six population schemes.

\begin{table}
\begin{center}
\caption{Best-fit parameters for the 3 galaxy classes with the 6 schemes for the model of constant
 GORES fraction per class}

\begin{tabular}{lllcccc}
\hline
 & \multicolumn{2}{c}{$r<R_{\rm v}$ AND} \\
\cline{2-3}
Scheme & $v_r < v_{r,\rm crit}$ & $v_r > -v_{r,\rm
  crit}$ & $f_{\rm v}$  & $f_{\rm i}$ & $f_{\rm b}$ & $\chi_\nu^2$ \\
\hline
0 & \ virial & \ virial     & 0.11 & 0.33 & 0.26 & 4.6 \\
1 & \ infall & \ virial     & 0.10 & 0.34 & 0.24 & 3.5 \\
2 & \ infall & \ backsplash & 0.10 & 0.33 & 0.28 & 2.9  \\
3 & \ infall & \ infall     & 0.10 & 0.34 & 0.23 & 2.7  \\
4 & \ backsplash & \ backsplash & 0.09 & 0.32 & 0.32 & 3.0 \\
5 & \ backsplash* & \ backsplash & 0.10 & 0.33 & 0.30 & 2.9 \\
\hline
\label{pops}
\end{tabular}
\end{center}
\smallskip
{\it Notes:} 
$\chi_\nu^2$ is the $\chi^2$ per degree of freedom. 
The errors are typically 0.01 on $f_{\rm v}$ and $f_{\rm i}$ and 0.04 on
$f_{\rm b}$. In Scheme~5, the extreme negative velocities for $r<R_{\rm v}$
 correspond to infall galaxies.
\medskip
\end{table}
 The best-fit predictions for the scheme with the lowest reduced $\chi^2$
 (hereafter $\chi_\nu^2$), Scheme~3
(where both the very positive and very negative radial velocity objects within the
 virial sphere are considered to be part of the 
infall population), provide a reasonable match to the observed fractions of GORES in bins of projected
 phase space. But $\chi_\nu^2 = 2.7$ for this best-fit model, suggests
 that the model itself can be improved.

\subsubsection{Model 2: Recent starburst fractions increasing with physical radius}
\label{model2}
 In our second model, we suppose that the fraction of GORES is no
 longer constant per class, but varies with physical radius $r$ as in
 Eq.~\ref{modelfit}, varying the normalisation and scale for each class:
\begin{equation}
f_\alpha(r) = f_\alpha {r/R_{\rm v}\over r/R_{\rm v} + a_\alpha} \ ,
\label{fofr}
\end{equation}
i.e. rising roughly linearly with radius for $r < a_\alpha$, and saturating to
an asymptotic value $f_\alpha$ at large radii.
The predicted fractions of GORES is then
\begin{equation}
g_{\rm GORES}(R_i,v_j) = \sum_\alpha p(\alpha|R_i,v_j) \sum_k f_\alpha(r_k) \,
q(r_k|R_i,v_j,\alpha), 
\label{gvary}
\end{equation}
where $q(k|R_i,v_j,\alpha)$ is the fraction of the particles of class $\alpha$ in
the cell of projected phase space $(R_i,v_j)$ that are in the $k$th bin of
 physical radius. The physical radial bins are expressed as $r_k\!=\!R_i\,\cosh u_k$,
 where $u_k$ is linearly spaced from 0 to $\cosh^{-1} r_{\rm max}/R_i$,
 using $r_{\rm max}\!=\!50\,R_{\rm v}$.

\begin{table*}
\begin{center}
\caption{Best-fit parameters for the three galaxy classes with the four schemes for models of
 radially-varying GORES fraction}
\tabcolsep 4.5pt
\begin{tabular}{lllcccccccrlclcl}
\hline
 & \multicolumn{2}{c}{$r<R_{\rm v}$ AND} \\
\cline{2-3}
Scheme & $v_r < v_{r,\rm crit}$ & $v_r > -v_{r,\rm
  crit}$ & $f_{\rm v}$  & $f_{\rm i}$ & $f_{\rm b}$ 
& $a_{\rm v}$ & $a_{\rm i}$ & $a_{\rm b}$ & 
$\chi_\nu^2$ & \multicolumn{1}{c}{$F$} & $P_F$ & $\Delta \rm AICc$ & \!\!\!$P_{\rm
  AICc}$  \\
\hline
0 & \ virial & \ virial     & 0.32 & 0.38 & 0.35 & 0.58 & 0.58 & 0.58 & 2.0 &
46 \ \ \  & 0 & 87 & 0\\
1 & \ infall & \ virial     & 0.24 & 0.37 & 0.29 & 0.36 & 0.36 & 0.36 & 1.8 &
31  \ \ \ & 0 & 55 & 0  \\
2 & \ infall & \ backsplash & 0.21 & 0.36 & 0.30 & 0.28 & 0.28 & 0.28 & 1.7 &
24  \ \ \ & 0 & 39 & 0\\
3 & \ infall & \ infall     & 0.19 & 0.36 & 0.26 & 0.22 & 0.22 & 0.22 & 1.6 &
24  \ \ \ & 0 & 36 & 0\\
4 & \ backsplash & \ backsplash & 0.23 & 0.35 & 0.35 & 0.35 & 0.35 & 0.35 &
1.8 & 24  \ \ \ & 0 & 40 & 0 \\
\cline{15-16}
5 & \ backsplash* & \ backsplash & 0.22 & 0.36 & 0.34  & 0.31 & 0.31 & 0.31 &
1.8 & 22 \ \ \  & 0 & 37 & 0 & AIC--AIC(3) & \multicolumn{1}{c}{$P$} \\
\hline
0 & \ virial & \ virial     & 0.41 & 0.35 & 1.00 & 0.88 & 0.27 & 4.79 & 2.0 &
0.6 & 0.56 & --3\ \ \ \ \ \  & \ \,--- & 11.8\ \ \  & 0.003 \\
1 & \ infall & \ virial     & 0.34 & 0.34 & 0.36 & 0.72 & 0.00 & 1.23 & 1.6 &
3.6 & 0.04 & 5.7 & 0.06 & 7.6 & 0.02 \\
2 & \ infall & \ backsplash & 0.30 & 0.33 & 0.21 & 0.62 & 0.00 & 0.00 & 1.5 &
3.8 & 0.03 & 5.5 & 0.06 & 4.4 & 0.11 \\
{\bf 3} & \ {\bf infall} & \ {\bf infall}     & {\bf 0.29} & {\bf 0.34} &
{\bf 0.39} & {\bf 0.62} & {\bf 0.00} & {\bf 1.40} & {\bf 1.2} & {\bf
5.7} & {\bf 0.01} & {\bf 8.3} & {\bf 0.02} & --- &
\multicolumn{1}{c}{---}\  \\
4 & \ backsplash & \ backsplash & 0.26 & 0.37 & 0.28 & 0.50 & 0.55 & 0.00 &
1.6 & 3.1 & 0.06 & 4.1 & 0.13 & 5.4 & 0.07 \\
5 & \ backsplash* & \ backsplash & 0.27 & 0.35 & 0.27 & 0.53 & 0.19 & 0.00 &
1.6 & 2.8 & 0.08 & 3.4 & 0.19 & 5.7 & 0.06 \\
\hline
\label{popswr}
\end{tabular}
\end{center}

\vbox{Notes: The upper portion is for fits in which $a_{\rm v}\!=\!a_{\rm
  i}\!=\!a_{\rm b}$ is enforced, while
the lower portion is for fits with fully free scale radii.
The errors are typically 0.01 on $f_{\rm v}$ and $f_{\rm i}$ and 0.04
on $f_{\rm b}$.
In Scheme~5, the extreme negative velocities for $r<R_{\rm v}$ are infall.
The best fitting scheme 3 is highlighted in bold.
$F$ and $P_F$ represent the $F$ test statistic that the fit with the scheme
is significantly better than that of the previous model ($f(r)$ with $a_{\rm
  v}\!=\!a_{\rm i}\!=\!a_{\rm b}$ vs. $f\!=\!\rm cst$ for the upper panel and general
$f(r)$ vs. $f(r)$ with $a_{\rm v}\!=\!a_{\rm i}\!=\!a_{\rm b}$ for the lower panel)
 and the associated probabilities of a larger value of $F$ appearing by chance.
$\Delta$AICc and $P_{\rm AICc}$ represent the same quantities with the
Akaike criterion (${\rm AIC}=\chi^2+2\,k$, for $k$ parameters, \citealp{Akaike74}) modified for
finite size samples (${\rm AICc}=\chi^2+2\,k\,n/(n\!-\!k\!-\!1)$, for $k$ parameters
and $n$ data points,
\citealp{HT89}).
The final two columns for the lower panel provide the Akaike statistic
testing whether the fit for $\rm Scheme \neq 3$ is as good as Scheme 3 and its
associated probability that a larger value of AICc occurs by chance.}
\end{table*}

\begin{figure}
\centering
\includegraphics[width=8.7cm]{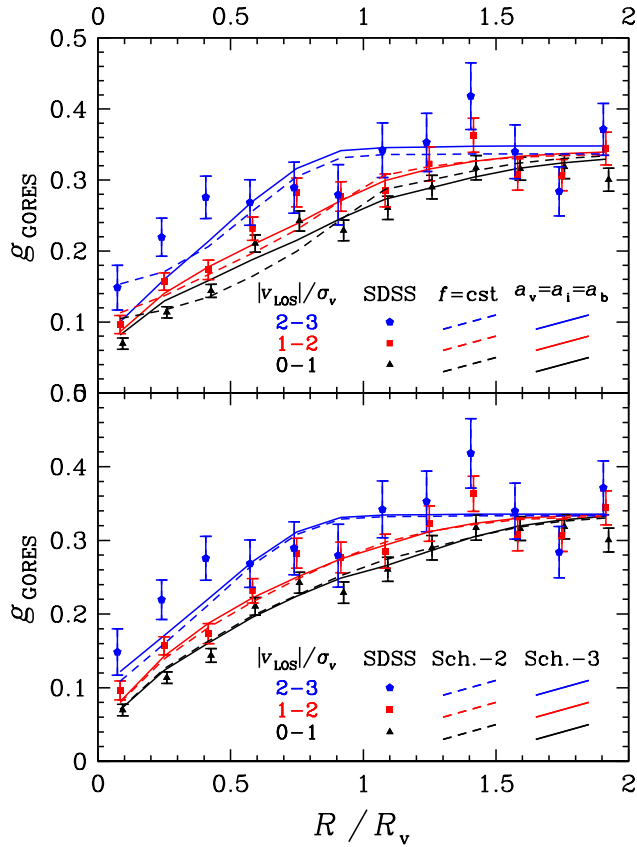}
\caption{Same as Fig.~\ref{fGORESvsR}, with best fit models overplotted: \emph{Top}: 
the \emph{dashed} and \emph{solid curves} show the best-fit Scheme~3 
models (Tables~\ref{pops} and \ref{popswr}) assuming
no radial dependence (using Eq.~\ref{gcst}) and the radially varying one of
Eq.\ref{fofr}, forcing $a_{\rm v}\!=\!a_{\rm i}\!=\!a_{\rm b}$, using
Eq.~\ref{gvary}, respectively. \emph{Bottom}: Best-fit radially varying
model (Eq.~\ref{fofr}) with free values of $a_{\rm v}$, $a_{\rm i}$, and
$a_{\rm b}$, for Schemes 2 (\emph{dashed}) and 3 (\emph{solid}).
The schemes are defined in Tables~\ref{pops} and \ref{popswr}.}
\label{fGORESvsRwmodels}
\end{figure}

Table~\ref{popswr} shows the new set of fits. The first set of lines shows
 the fits when we force the same quenching radius for the three 
classes: $a_{\rm v}\!=\!a_{\rm i}\!=\!a_{\rm b}$.
Again, Scheme 3 provides the lowest $\chi_\nu^2$, which is now 1.6 (in
comparison with 2.7 for the $f\!=\!\rm constant$ model). So with the inclusion of
 just one extra parameter, the radially-varying model fits the data much better.
 This is confirmed with both the $F$-test and the AICc \citep{Akaike74,HT89} 
criterion (Table~\ref{popswr}). Note that cosmic variance on the finite number of
 clusters has been incorporated in the errors.
The value of quenching radius $a$ is low in Scheme 3, which indicates that it is not
that far from the model with constant fraction of GORES per class. The top panel of 
Fig.~\ref{fGORESvsRwmodels} shows the observed fractions of GORES together with the
 best fits with Scheme 3 for both the constant fraction per class and radially-varying
 models with  $a_{\rm v}\!=\!a_{\rm i}\!=\!a_{\rm b}$.
 Both models capture the initial rise of the GORES fraction with projected radius 
 and subsequent saturation, as well as the velocity modulation, even though
 the fit is not excellent.
 
In the next set of fits, shown in the bottom six rows of Table~\ref{popswr}, we
lift the assumption of $a_{\rm v}\!=\!a_{\rm i} = a_{\rm b}$.
The best-fit reduced $\chi^2$ values are now $\leq 1.6$ for Schemes 1--5, with an
often significant improvement of the fit (despite the two extra parameters) over 
the radially-varying model with the same scale parameters (see the $F$ test
statistic in Table~\ref{popswr}). The $F$-test indicates that the improvement
 is significant with the inclusion of two extra parameters (three scales
 instead of a unique one). Using the Akaike criterion, which reduces here to
 $\Delta {\rm AICc}=\Delta{\rm AIC}=\Delta \chi^2$, with associated probability
 $P=\exp(-\Delta{\rm AIC}/2)$, we conclude that
Scheme~3, which still leads to the lowest $\chi_\nu^2=1.2$, is
a significantly better fit  than Scheme~0 (99.7\% confidence) and
marginally significantly better fit than Schemes 4 (93\% confidence) and 5
(94\% confidence).
Although a decent fit to the data, this best-fit model still predicts too
low values of $f_{\rm GORES}$ for high absolute LOS velocity galaxies within
half virial radius of the clusters.
Interestingly, the best-fit model (Scheme~3) predicts a constant GORES
fraction for the infall class galaxies ($a_{\rm i} = 0$), while the next
best-fitting model (Scheme~2) predicts that both infall and backsplash
populations have a constant GORES fraction ($a_{\rm i} = a_{\rm b} = 0$).

 \section{Discussion}
 \label{discussion}
  
 This paper is a unique analytical effort to explore the variations in
 the spectral and physical properties of different dynamical classes
 of galaxies residing in similar environments projected on the
 observed sky (projected radius and absolute LOS velocity), 
 and to interpret these differences.  We draw a
 sample of $\sim\!20,000$ galaxies ($M_{r}\!\leq\!-20.5$) found within
 2\rv and $|\Delta v_{\rm LOS}| < 3\,\sigma_v$ of 268 galaxy clusters, and a
 similar sample of field galaxies for this purpose. All the galaxies
 are taken from the SDSS DR4 spectroscopic galaxy catalogue.

 \subsection{Velocity segregation of stellar mass}
 \label{masseg}

Three physical mechanisms might explain the excess of high mass
galaxies in the cluster cores amongst the low velocity galaxies
 (Fig.~\ref{frachimass}): cluster tides, two-body relaxation, and dynamical
friction. We examine in turn each one of these.

Tidal effects from the cluster gravitational field (predicted by
\citealp*{merritt83,Mamon92,GHO99}, clearly seen in cosmological
simulations 
by \citealp{Ghigna+98}, and measured through
gravitational lensing by \citealp{Natarajan+09}) will be most effective on slowly
 moving galaxies, because these have more time to feel the cluster tides. 
Even though stars are less affected by tides than the more extended dark matter
haloes, 
simulations of a live dwarf spiral galaxy orbiting  a spiral  galaxy $\sim$100
times more massive, with fixed gravitational potential
\citep{Klimentowski+09} or live $N$-body system 
\citep{Lokas+10} indicate that, at each pericentric
passage, over 35\% (\citeauthor{Klimentowski+09}) to 67\% (\citeauthor{Lokas+10})
of the stars of a spiral galaxy  are tidally stripped. 
Tides will thus transform massive galaxies into low mass ones, and low mass galaxies into
even lower mass ones. So tides cause a decrease in the fraction of high mass
galaxies, which disagrees with the trend seen in Fig.~\ref{frachimass}.

If two-body relaxation can lead to energy equipartition in the
cluster cores, then the high mass galaxies will move slower (in 3D). So the
fraction of high mass galaxies among the low 3D velocity ones should increase
with decreasing cluster-centric radius. Hence, the fraction of high mass
galaxies among the low absolute LOS velocity should also increase
with decreasing radius. However, the same argument would lead
us to predict that the fraction of high mass galaxies among the high velocity
ones should decrease with decreasing radius, and instead Fig.~\ref{frachimass}
shows no such trend. Therefore, the trend in Fig.~\ref{frachimass}
cannot be due to two-body relaxation, unless some other process is increasing
the fraction of high mass galaxies among the high velocity ones within the
cluster core.

The time required by dynamical friction to make the orbits of galaxies
decay towards the cluster centre scales as $v^3/m$, where $v$ is the 3D
velocity of the galaxy of mass $m$. Therefore the most massive and slowly-moving 
galaxies should have their orbits decay the most, and be brought to even
 lower (projected) radius. So dynamical friction should
boost the fraction of high-mass slow-moving galaxies in the cluster core.
Moreover, the effects of dynamical friction should be weak for the rapidly
moving galaxies.
 This explains the lack of trend in the fraction of high mass
galaxies with projected radius for the galaxies with large absolute
LOS velocities.

For dynamical friction (and tides), the high-velocity galaxies act little, if
one assumes a population in virial equilibrium. Now, the infalling galaxies
also tend to have high absolute radial velocities (Fig,~\ref{rvr}), hence
high absolute LOS velocities (e.g., Table~\ref{tabfracs}),
in fact even higher than the virialised population. So the
same argument applies to the infalling galaxies: their velocities are too
high to be seriously affected by dynamical friction or tides.

We note that the two-body relaxation timescale is independent of galaxy mass $m$,
while the dynamical friction timescale varies as $1/m$. It is easy to show
that the ratio of two-body relaxation time to dynamical friction time scales
as $\langle m \rangle/m$. Thus, the lack of  high-velocity low mass galaxies in
the inner bin of projected radii in Fig.~\ref{frachimass}, in contrast with
the excess expected from two-body relaxation, is not inconsistent
with the excess of high-mass galaxies at low radii observed in
Figure~\ref{frachimass} and expected from dynamical friction.

Is dynamical friction efficient enough to increase the fraction of high mass
galaxies towards the centre among the slowly moving galaxies?
We adopt the dynamical friction timescale that \cite{Jiang+08} carefully
calibrated with a cosmological hydrodynamical simulation: 
\begin{equation}
t_{\rm df} \simeq 1.4\, {M(r)/ m\over \ln [1+M(r)/m]}\,{r\over v_{\rm
    c}(r)} \, ,
\label{tdf}
\end{equation}
where $m$ is the galaxy mass (with its tidally stripped DM halo) and
$v_c(r) = \sqrt{GM(r)/r}$ is the circular velocity at radius $r$.
Eq.~\ref{tdf} incorporates the effects of elongated orbits and tidal
stripping.
\begin{figure}
\centering
\includegraphics[width=\hsize]{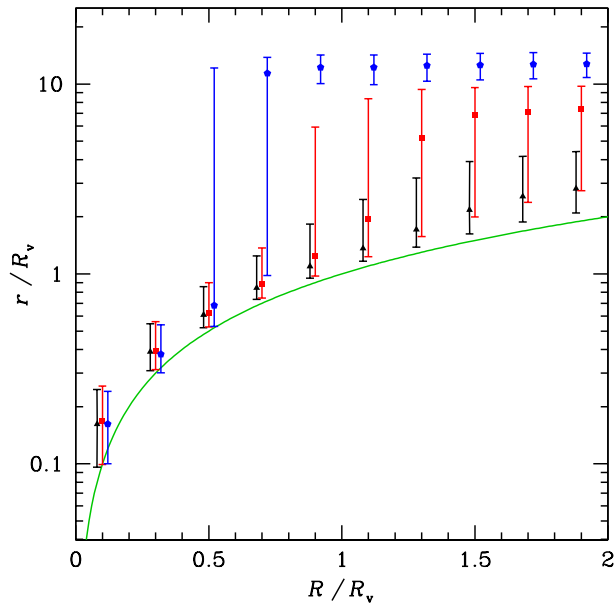}
\caption{Physical radius versus projected radius for particles of stacked
 mock cluster (medians with error bars for quartiles) in bins of absolute LOS velocity:
0--1$\,\sigma_v$ (\emph{black triangles}),
1--2$\,\sigma_v$ (\emph{red squares}), and 
2--3$\,\sigma_v$ (\emph{blue pentagons}).  
The \emph{green curve} shows $r=R$.
\label{rvsR}}
\end{figure}
One would like to have an idea of the physical radius of each galaxy given
its projected radius and absolute LOS velocity.
Fig.~\ref{rvsR} shows the distribution of physical radii given the projected
radii in our adopted bins of absolute LOS velocity.
 \emph{The physical radii of the low absolute LOS velocity galaxies are
 close to their projected radii}, so that these low-velocity galaxies can be
 used to deproject radial trends. But high absolute LOS velocity galaxies
 appear to systematically lie very far from the virial sphere (typically
 10\,$R_{\rm v}$).
\begin{figure}
\centering
\includegraphics[width=\hsize]{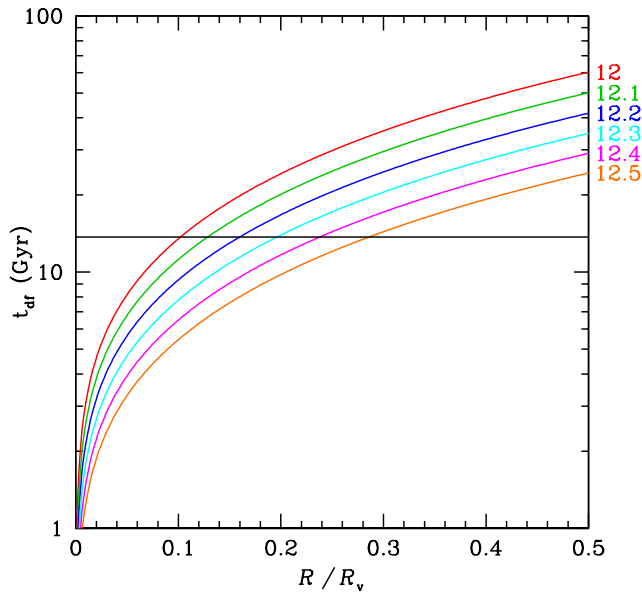}
\caption{Timescale for orbital decay by dynamical friction versus projected
 radius for galaxies of log total mass  (in solar masses, including the tidally stripped dark
 matter halo) of (from top to bottom, labelled on right)
12 (\emph{red}),
12.1 (\emph{green}),
12.2 (\emph{blue}),
12.3 (\emph{cyan}),
12.4 (\emph{magenta}),
and 12.5 (\emph{orange}),
  orbiting an NFW cluster of scale radius $0.19\,R_{\rm v}$ and mass
  $M_{\rm V} \equiv M_{100} = 1.9\times10^{14}{\rm M_\odot}$ (corresponding
 to the median mass of $M_{180}=10^{14.2}\,{\rm M_\odot}$ of the SDSS clusters
 studied here) within the virial radius $R_{\rm v}\equiv R_{100} =1.5 \,\rm Mpc$.
The timescale is estimated using Eq.~\ref{tdf} assuming that
$r=1.3\,R$ as inferred from our cosmological simulation (see
Fig.~\ref{rvsR}). The age of the Universe is shown as the \emph{horizontal line}.
\label{tfricvsR}}
\end{figure}

Using the relation $r\approx 1.3 R$ from Fig.~\ref{rvsR}, we can estimate the
dynamical friction time as a function of projected radius.
Fig.~\ref{tfricvsR} indicates that a galaxy (along with its dark halo)
must have a total mass of at least $10^{12.45} {\rm M_\odot}$, for the
orbital decay time, at $R=0.25\,R_{\rm v}$, to be smaller than the age
of the Universe. In comparison, \citeauthor{YMvdB09}
(\citeyear{YMvdB09}, their table 4) found that in the SDSS, central
group/cluster galaxies of mass $3\times 10^{10} {\rm M_\odot}$ resides
in halos of mass $10^{12.2} {\rm M_\odot}$.  Also, according to
table~4 of \citeauthor{YMvdB09}, central haloes of mass $10^{12.45}
{\rm M_\odot}$ have corresponding galaxy masses of $10^{10.66} {\rm
  M_\odot}$, i.e. 45\% above our galaxy mass threshold.  We therefore
conclude that dynamical friction might indeed be just sufficiently
effective to explain the trends in Fig.~\ref{frachimass}. However, the
DM halo of a galaxy extends further out than its stellar component, so
halos are tidally stripped in a more efficient manner (e.g.,
\citealp{Klimentowski+09}). This means that the ratio of total to
stellar mass of cluster galaxies is lower than for field galaxies. One
would therefore expect  the actual halo masses to be lower than
inferred for the central galaxies, from table~4 of \cite{YMvdB09}, and
thus the orbital decay times by dynamical friction to be somewhat longer.

 \subsection{The star-forming galaxies}

The distributions of specific star formation rate, broad-band colour, and
amplitude of the $4000\,\rm \,\AA$ break show statistical differences with absolute
LOS velocity at almost all radii out to $R<1.5\,R_{\rm v}$
 (Figs.~\ref{gr}, \ref{ssfr}, and \ref{d4}; 
 Tables~\ref{table:gr}, \ref{table:ssfr}, and \ref{table:d4}, respectively).
However, the field component shows a bimodal distribution in
 \ehd\ (Fig.~\ref{hd}) and D$_n$4000 (Fig.~\ref{d4}) around 2\,\AA~and 1.5 respectively.
 This is in agreement with the bimodality of these two diagnostics when plotted in terms of
 stellar mass, luminosity, or concentration \citep{kauff03a}. We
 adopt these thresholds to define a galaxy as a recent starburst galaxy
 (GORES) if it has \ehd\,$>\,2$\,\AA\ and D$_n$4000\,$<\!1.5$. 
 
As expected, the fraction of GORES drops within the
 virial radius \rv for all velocity classes (Fig.~\ref{fGORESvsR}). 
But \emph{the fraction of GORES is modulated by the absolute LOS velocity: at a 
  given projected radius, galaxies with higher  $|v_{\rm LOS}|$ have higher
 fractions of GORES.} For the high absolute LOS velocity galaxies, the fraction of
 star-forming galaxies declines rapidly but only well within the core region
 ($\lesssim\!0.5\,R_{\rm v}$). This once again strengthens our
 interpretation that the high-velocity galaxies are the youngest
 members of the core, having recently fallen in.

A curious feature occurs in the radial variation of both
high-mass (Fig.~\ref{frachimass}) and GORES fractions (Fig.~\ref{fGORESvsR}): the galaxies with
 high absolute LOS velocity show a marginally significant peak (dip) in
 the fraction of GORES (high mass galaxies) at 1.5$\,R_{\rm v}$.
If this feature is not a statistical fluke, it may be reflecting
 the discontinuity seen in SFR of cluster galaxies at the physical radius
 where infalling and virialised galaxies meet. However, this is expected to
 occur closer to the virial radius and our simple models involving the
 virial, infall and backsplash populations do not lead to any discontinuity
 of $f_{\rm GORES}$ at $1.5\,R_{\rm v}$ (Fig.~\ref{fGORESvsRwmodels}).
An explanation may require a model going beyond spherical symmetry, e.g.,
with large-scale filaments feeding clusters
 \citep[see][]{porter07,porter08,mahajan11}. For the benefit of the reader,
 we emphasise that the enhancement seen in the SFR of galaxies seems to occur in a
 very narrow region only. This observation is very
 interesting and still being explored. It might prove useful to distinguish
 between the backsplash and infall galaxies.
 
Given the $M^*$-SSFR anti-correlation (\eg\ \citealp*{MPP01}; \citealp{noeske07b,damen09}),
 the peak in the fraction of low mass galaxies might be directly linked to
 the enhancement of SFR in infalling galaxies  
 on the outskirts of clusters before the environmental effects in
 clusters predominate \citep{porter07,porter08}.

While the projected radii of the low absolute LOS velocity particles in the
stacked mock cluster  underestimate
only slightly their physical radii (Fig.~\ref{rvsR}), our cosmological
simulation indicates that the high absolute LOS velocity particles with
projected radii $R > 0.8\,R_{\rm v}$ have very
large physical radii (typically over 10 virial radii). This implies that the
observed fraction of GORES for the high absolute LOS velocity bin at $R >
R_{\rm v}$ should be the same as that of the field, but this is not what is
observed (Fig.~\ref{fGORESvsR}). One cannot resort to cosmic variance to
explain  this discrepancy, given the large number of clusters used
in our analysis of the SDSS observations (268) and in the simulations (93),
and we cannot find any other convincing explanation for this dichotomy.

\subsection{Identifying backsplash galaxies}
\label{bspid}

The analysis of the $z\!=\!0$ output of DM particles in our cosmological hydrodynamical
simulation allows us to distinguish backsplash particles, whose positions
in radial phase space imply that they previously crossed their parent
(mock) clusters from virialised particles (within the virial sphere) and the
remaining infall particles (which are actually expanding away from their
clusters when they are beyond the turnaround radius of $\simeq 3.5\,R_{\rm v}$).
Our analysis in projected phase space indicates that the best place to detect
backsplash particles is just beyond the virial radius and at very low velocities.
With Scheme~3, the backsplash fraction reaches 54\% (bottom panel of
Fig.~\ref{bspfrac}) for the lowest absolute LOS velocity bin. 
With the maximum backsplash Scheme~5, it reaches 75\% for the lowest
 $|v_{\rm LOS}|$ bin.

\citet{Pimbblet11} compared the distribution of SSFRs of galaxies with $0.3 <
|v_{\rm LOS}|/\sigma_v < 0.5$, attributed to the infall class with the full
distribution. This led him to predict that the fraction of backsplash
galaxies decreases as $f_{\rm b} = 0.61 - 0.052\,R/R_{\rm v}$. Such a trend
(grey line in the bottom panel of Fig.~\ref{bspfrac})
appears inconsistent with the backsplash fractions predicted from our
cosmological simulation, even with our maximum backsplash Scheme~5. It is
difficult to understand how one would find that as many as half of all galaxies at 2
virial radii (the maximum projected radius in the analysis of
\citeauthor{Pimbblet11})
from clusters are backsplash. Backsplash galaxies typically only 
bounce out to 1--2 virial radii \citep{mamon04}.

Yet, the high fraction of backsplash galaxies found by
\citeauthor{Pimbblet11} agrees with the 54$\pm$20\% fraction of
backsplash particles that \citet{gill05} found in the range of projected
radii $1 < R/R_{\rm v}<2$ from their cosmological
simulations. In our Schemes 0--4, we find that only 17\%$\pm$1\%\footnote{The errors in our
fractions of backsplash particles are from 50 bootstraps on the 93 halos of the cosmological simulation.}
 of the particles
in this range of projected radius (with $|v_{\rm LOS}|<3\,\sigma_v$)
are backsplash particles. However, our
backsplash fraction amongst DM particles is
marginally consistent with the higher fraction found by \citeauthor{gill05} 
Even in the maximum backsplash Scheme~5 (partly motivated by \citeauthor{Pimbblet11}'s high
backsplash fraction, but only marginally consistent with the observed
fraction of GORES), we find only 40$\pm$1\% of the particles in the same
range of projected radius are backsplash. This is apparently inconsistent with the
fraction derived by \citeauthor{Pimbblet11}, although fully consistent
with that from \citeauthor{gill05}'s simulations. 

 Part of our disagreement with \citeauthor{Pimbblet11} on the fraction of
 backsplash galaxies in the range $1 < R/R_{\rm v}<2$ may be caused by
 the narrow range of $|v_{\rm   LOS}|/\sigma_v$ used by him to select the infall galaxies.

 \subsection{The quenching of star formation in 3D: infalling vs backsplash galaxies}
  
 The SDSS spectroscopy provides the means to address the role of the dynamical classes
 (virial, infall and backsplash) in shaping the properties of galaxies by studying the
 spectral features as a function of both projected radius \emph{and} absolute LOS velocity.

 Beyond the well-known radial dependence of galaxy properties in
 clusters, our analysis highlights the velocity modulation of these
 trends for the parameters 
 SFR/$M^*$, \ehd\ and the D$_n$4000. All trends with
 projected radius are amplified with increasing absolute LOS
 velocity. As can be seen in Fig.~\ref{rvsR}, the effect of $|v_{\rm
   LOS}|$ effectively adding to the projected radius is not trivial.
 Nevertheless, as qualitatively expected from the Hubble flow, in
 every bin of projected radius the higher absolute LOS velocities
 leads to higher physical radius.

In \S\ref{analysis} we find observational evidence to support the
statistical differences between the low velocity
($|v|_{\rm LOS}\!\leq\!\sigma_{v}$) and the high velocity
($2\!<\!|v_{\rm LOS}|/\sigma_{v}\!\leq\!3$) galaxies on the outskirts of 
clusters (1--1.5$\,R_{\rm v}$). K-S tests between the
different distributions in the clusters and in the field show
statistically significant differences between the low and high
velocity galaxies between 1--1.5\rv in SFR/$M^*$, \ehd~and the
D$_n$4000 (Tables~\ref{table:lgm}-\ref{table:d4}).

Despite the noisiness of the variations with projected radius of the numbers
of galaxies and of GORES, we have been able to deproject the fraction of
GORES to show that it is well fit (Fig.~\ref{deproj})
by a saturated linear model (Eq.~\ref{modelfit}).
Such a model applied in turn to the three different classes of galaxy
populations (virialised, infalling and backsplash) with the same quenching
radius 
($a_{\rm v}\!=\!a_{\rm i}\!=\!a_{\rm b}$), provides an adequate fit to the
 variation of GORES with projected radius in three bins of absolute LOS velocity
 (top four rows of Table~\ref{popswr} and upper panel of Fig.~\ref{fGORESvsRwmodels}).
 The model reproduces the rise with projected radius and the higher fractions
 for higher absolute LOS velocities. Similar models with free star formation
 quenching radii fit the data even better (bottom four rows of Table~\ref{popswr}
 and lower panel of Fig.~\ref{fGORESvsRwmodels}).
 In particular, the radius-dependent model where galaxies within the
 virial sphere ($r\!<\!R_{\rm v}$)
 with radial velocity $|v_r|\!>\!|v_{r,\rm crit}|$ are considered to
 be infalling, provides the best fit to the data (\S\ref{model12},  
 Fig.~\ref{rvr} and Table~\ref{popswr}).

In contrast, a simpler model, where the fraction
of GORES is a constant that only depends on the class of galaxy population,
does not provide an adequate fit (Table~\ref{pops}), 
even if it also reproduces these
qualitative trends (Fig.~\ref{fGORESvsRwmodels}).

 The ability of Scheme~3 to better fit the SDSS data than the other three 
 Schemes suggests that the high absolute \emph{radial} 
velocity galaxies within the virial sphere,
 both infalling and outflowing, retain their fraction of GORES. However, our
 best-fit model with Scheme~2 is also an adequate fit, so the model with
 quenching at pericentre is also consistent with the data.

In the constant GORES
 fraction model with Scheme~3, the GORES fraction in the backsplash class 
 is half-way between the analogous values for the virial and infalling populations,
 suggesting that one passage through the cluster environment causes half the quenching
 of star formation. We  now consider the deprojected models to assess how
 star formation is quenched at a unique radius where the three classes can be
 compared: the virial radius.

\begin{figure}
\centering
\includegraphics[width=8.7cm]{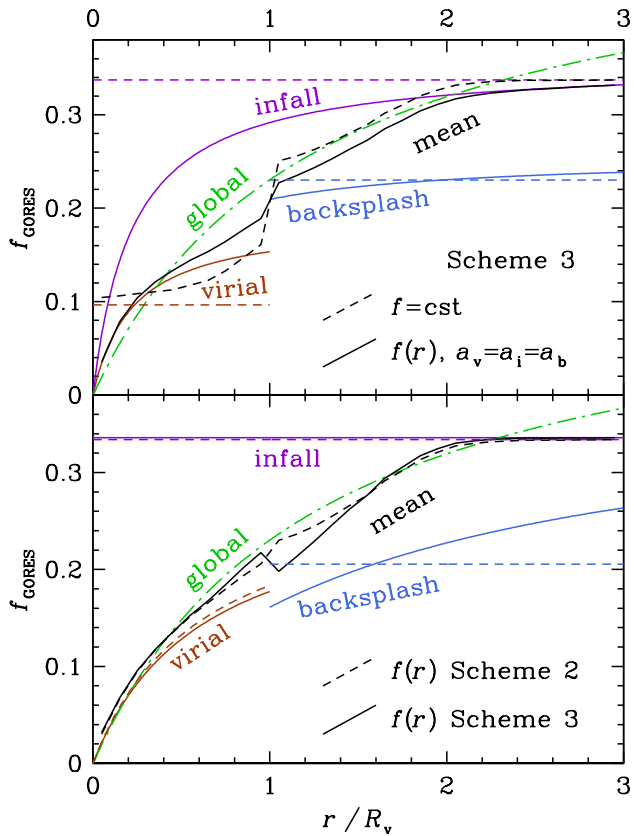}
\caption{Fraction $f(r)$ of GORES versus physical radius, 
for the virial (\emph{brown}), infall (\emph{purple}) and backsplash (\emph{blue-grey})
populations, their mean (using the cosmological simulation to determine the
radial variation of the densities of the three classes, \emph{black}) and the
global deprojection (\emph{green dash-dotted curve}).
\emph{Top}: best fit constant (\emph{dashed lines}) and radially increasing
(Eq.~\ref{modelfit}) 
with $a_{\rm v}\!=\!a_{\rm
  i}\!=\!a_{\rm b}$ (\emph{solid curves}), both with Scheme~3.
\emph{Bottom}: best fit radially increasing (Eq.~\ref{modelfit}) with free
star formation quenching radii for Schemes~2 (\emph{dashed}) and 3
(\emph{solid curves}).
}
\label{fsbvsr}
\end{figure}

 Fig.~\ref{fsbvsr} shows how the best-fitting models predict the fraction of
 GORES as a function of the 3D radius.
We first note that for the constant $f_{\rm GORES}$ per class model,
the mean variation of $f_{\rm GORES}$ (top panel of Fig.~\ref{fsbvsr})
departs more strongly from the global trend found
by the global deprojection (\S\ref{deprojsec}) than do the mean variations with
the other models. This reinforces the view that the constant $f_{\rm GORES}$
per class model does not reproduce well our data.

With the radially-dependant model of the GORES fraction with equal quenching
scale radii, again with Scheme~3, the fraction of GORES in the backsplash
 population at $r=R_{\rm v}$ is intermediate between those of the infalling
 and virialised populations, although closer to the virialised one (top panel
 of Fig.~\ref{fsbvsr}).
\begin{table}
\begin{center}
\caption{Quenching at $r=R_{\rm v}$ (models with free quenching
  radii)
\label{tabquench}
}
\tabcolsep 3pt
\begin{tabular}{lcccc}
\hline
Scheme & $f_{\rm GORES}^{\rm vir}(R_{\rm v})$ & $f_{\rm GORES}^{\rm inf}(R_{\rm
  v})$ & $f_{\rm GORES}^{\rm bsp}(R_{\rm v})$ & quenching \\
\hline
0 &  0.22 &  0.28 &  0.17 &  1.7 \\ 
1 &  0.19 &  0.34 &  0.16 &  1.3 \\ 
2 &  0.18 &  0.33 &  0.21 &  0.9 \\ 
{\bf 3} & {\bf  0.18} & {\bf  0.34} & {\bf  0.16} & {\bf  1.1} \\ 
4 &  0.17 &  0.24 &  0.28 & --0.7\ \ \,\\ 
5 &  0.18 &  0.30 &  0.27 &  0.2 \\ 
\hline
\end{tabular}
\end{center}

Note: the last column lists $Q=(f_{\rm GORES}^{\rm inf}-f_{\rm GORES}^{\rm
  bsp})/(f_{\rm GORES}^{\rm inf}-f_{\rm GORES}^{\rm vir})$, which measures the
effectiveness of quenching of the backsplash population relative to the
infall and virial ones.
The best fitting scheme 3 is highlighted in bold.
\end{table}

Table~\ref{tabquench} displays the fraction of GORES at the surface of the
virial sphere ($r=R_{\rm v}$) for the better-fitting radially-varying models
 with free quenching radii.
The backsplash quenching factor, $Q=(f_{\rm GORES}^{\rm inf}-f_{\rm GORES}^{\rm
  bsp})/(f_{\rm GORES}^{\rm inf}-f_{\rm GORES}^{\rm vir})$, should lie between
0 (no quenching: backsplash and infall have the same GORES fraction at a given
radius) and 1 (full quenching: backsplash and virial have same GORES fraction
at a given $r$). The values above unity imply that the fraction of GORES in the
backsplash population lying on the virial sphere is lower than the
corresponding fraction of the virial population at the same location. This
leads to the unphysical result that on the virial sphere backsplash
galaxies are more passive than the virialised ones.

For our best-fitting Scheme~3, $Q\!=\!1.1$ is just above unity, i.e. the
backsplash and virialised populations on the virial sphere 
have very similar fractions of GORES\footnote{Forcing $Q\!=\!1$ in our fit leads
  to $\chi_\nu^2$ only 0.003 higher for Scheme~3.}, while for the second best
fitting Scheme~2, $Q\!=\!0.9$ (see bottom panel of Fig.~\ref{fsbvsr}).
On the other hand, for Scheme~4, $Q$ is much smaller than 0: the fraction of
GORES for galaxies lying on the virial sphere is higher for the backsplash
galaxies than for the infall population. This again appears to be an
unphysical result, which leads us to disregard Scheme 4 (which fits the data
 worse than Scheme 3, but with only marginal significance; see Table~\ref{popswr}).
This suggests that \emph{the backsplash
 galaxies are strongly quenched relative to the infalling ones}, at least
half the way from infalling to virialised galaxies at $r=R_{\rm v}$
(for Scheme~5, which is only marginally consistent with the data) to nearly fully 
 quenched to the level of virialised galaxies (Schemes 3 and 2).

In contrast, using hydrodynamical cosmological simulations of the
Local Group, \citet{knebe+11} find that the luminosity function of
 backsplash galaxies is similar to that of the infalling galaxies and
 that the total mass within the
radius containing the outermost stars divided by the luminosity is
higher for infalling galaxies than for backsplash galaxies. The first
result suggests little quenching of star formation as dwarf galaxies
cross through the virial sphere of the Milky Way, while the second
result suggests inverse quenching, that is that backsplash galaxies
have higher luminosity per unit mass. 
However, star formation in $10^{12} M_\odot$ galaxies falling into  
$10^{14} M_\odot$ clusters should be similar to the quenching of
star formation in dwarf galaxies of mass $10^{10} M_\odot$ falling
into the Milky Way virial sphere. What is more surprising is that
fig.~5 of \citet{knebe+11} indicates that galaxies in the
virialised class have higher luminosity for given internal velocity
dispersion than galaxies in the infall class. Analysing the residuals
of luminosity versus velocity dispersion relative to the mean global
trend, we deduce that the higher luminosity of virialised galaxies
compared to infall galaxies of the same velocity dispersion is
statistically significant (with 99.9\% confidence using a K-S test, and
 by typically two magnitudes). The backsplash population on the other hand
 is intermediate but not inconsistent with either the virial or the infall populations.
 The higher luminosities of the virial population of simulated dwarfs appear
 to contradict the observed positive luminosity -- radius correlation found
 for luminous virialised galaxies with cluster-centric distances above
 $\ga 0.1 R_{\rm v}$ from their parent rich clusters \citep{ABM98}.

The quest for the physical mechanism responsible for this quenching is
beyond the scope of this paper, be it tidal stripping, ram pressure
stripping, harassment, starvation or pre-processing in groups.
For instance, \citet{vandenBosch+08} argue that the
lack of dependence of galaxy colour on cluster mass is evidence against the
dominance of  ram pressure stripping and harassment.

We note that in our two best-fitting models (Schemes~3 and 2),
the fraction of GORES in infalling galaxies is independent of their physical
distance to the cluster. Our best-fit model requires a scheme (like our Scheme~3)
 where galaxies within the virial
sphere with high outgoing velocities are equivalent to the infalling
galaxies within the same volume (those with strong negative
velocities). This together with the comparison of GORES fraction at
the virial radius suggest that \emph{the star formation in galaxies is
  quenched on a timescale comparable to the time taken for a galaxy to
  bounce out of a cluster and reach its virial radius.}  This
timescale corresponds to about 1 to 2 Gyr, which is very close to the
lookback time to which our joint H$\delta$-D$_n$4000 GORES diagnostic
is most sensitive (\S\ref{sfdiag}). Correcting for this
lookback time, one concludes that \emph{star formation in a galaxy is
 nearly fully quenched in a single passage through the cluster}.

Of course, there will be a variety of orbital pericentres for the infalling
galaxies, leading to different degrees of quenching. But the typical high
elongations found for orbits in $\Lambda$CDM simulations \citep{Ghigna+98}
suggests that orbits with large pericentres will be rare.

We find that $17.9\!\pm\!0.4\%$\footnote{The errors here and below on the
 fractions of GORES are binomial and do not incorporate cosmic variance.}
of the galaxies within the \emph{virial cylinder} (projected radii
 $R\!<\!R_{\rm v}$ and $|v_{\rm LOS}| < 3\,\sigma_v$) are GORES. 
This can be compared to the fraction of blue galaxies within the virial
cylinder: we find that $20.3\pm0.4\%$ of the galaxies with $R<R_{\rm v}$ are 0.2
 magnitude bluer than the Red Sequence (this fraction is fairly consistent with the
 $\sim$78\% of  of SDSS cluster galaxies that \citealp{yang08} find  to lie on the
 Red Sequence). So, \emph{within the virial cylinder, the fraction
 of GORES is roughly 88$\pm$0.5\% of the fraction of blue galaxies}, although
 some blue galaxies are not GORES and some GORES are not blue \citep{mahajan09}.

In comparison, using the same cosmological simulation as in the present article, 
\citet{mbm10} recently showed
 that $23\!\pm\!1\%$ of galaxies within the virial cylinder are
 outside the virial sphere. As discussed by \citeauthor{mbm10}, the
 match between fraction of blue galaxies and fraction of cluster
 interlopers is probably a coincidence since some galaxies within the
 cluster sphere must be blue and some interlopers must be Red Sequence
 galaxies in projected groups.

With our models, we can estimate
 the fraction of GORES within the \emph{virial sphere} using the average of the mean
 fractions $f(r)$ (Fig.~\ref{fsbvsr}) weighted by $r^2$ times the galaxy
 number density profile, $\nu(r)$ measured in our cosmological simulation. 
Results given in Table~\ref{deprojfracs} indicate that for
our three best-fitting models (see Table~\ref{popswr}), \emph{the fraction of
GORES within the virial sphere is $13\pm1\%$}. We then predict the fraction of
 blue galaxies within the virial sphere to be $13/0.88 = 15\pm2\%$.
\begin{table}
\caption{Fractions of Galaxies with Ongoing or Recent Efficient Star Formation}
\begin{center}
\tabcolsep 3pt
\begin{tabular}{lcccccc}
\hline
Model & Scheme & Virial& Virial & Virial & Infall & Backsplash\\
& & cylinder & sphere & class & class & class \\
Range & & $R$$<$$R_{\rm v}$ & $r$$<$$R_{\rm v}$ & ($r$$<$$R_{\rm v}$) & $R$$<$$2\,R_{\rm v}$ &
$R$$<$$2\,R_{\rm v}$ \\
\hline
$f$=cst & 3 & 0.17 & 0.12 & 0.10 & 0.34 & 0.23 \\
$a_{\rm v}$=$a_{\rm i}$=$a_{\rm b}$ & 3 & 0.17 & 0.13 & 0.11 & 0.33 & 0.22 \\
$f(r)$ & 1 & 0.17 & 0.13 & 0.12 & 0.34 & 0.19 \\
$f(r)$ & 2 & 0.17 & 0.13 & 0.11 & 0.33 & 0.21\\
$\boldsymbol{f}\mathbf{(}\boldsymbol{r}\mathbf{)}$ & {\bf 3} & {\bf 0.17} & {\bf
    0.13} & {\bf 0.11} & {\bf 0.34} & {\bf 0.19}\\ 
$f(r)$ & 4 & 0.17 & 0.13 & 0.11 & 0.32 & 0.26 \\
$f(r)$ & 5 & 0.17 & 0.13 & 0.11 & 0.34 & 0.27 \\
\hline
SDSS & --- & \ \,0.176 & --- & --- & --- & --- \\
\hline
\end{tabular} 
\label{deprojfracs}
\end{center} 

Notes: The last six models are our best-fitting ones ($\chi_\nu^2 \leq
1.6$). 
The best fitting scheme 3 is highlighted in bold.
Note that the fractions of GORES within the virial cylinder slightly
underestimate the observed fraction, but are consistent with it within the 1\% errors.
\end{table}

Now, while most galaxies within the virial sphere are part of the virial population, 
some are infalling (Schemes~1 to 3) or backsplash (Scheme~2). Yet,
we can proceed even further and estimate the fraction of GORES among the
virialised class, again using our cosmological simulation to perform the
calibration.
As seen in Table~\ref{deprojfracs}, \emph{the fraction of GORES among the virialised
class is not zero, but typically 11$\pm$1\%}. We also predict that the
fraction of blue galaxies among the virialised population is $11/0.88 =
12.5\pm2\%$. Moreover, Table~\ref{deprojfracs} indicates that 
the fraction of GORES among the $R<2\,R_{\rm v}$ infall
galaxies is $34\pm1\%$ while among the backsplash galaxies the fraction of
GORES is typically $20\pm4\%$ (for Schemes~3, 2 and 1) or perhaps as high as $27\pm4\%$
(Schemes~4 and 5). 

The presence of GORES among
the virialised class might be caused by those low metallicity galaxies for
which our H$\delta$-D$_n$4000 estimator of recent efficient star formation
has a lookback time as long as 3 Gyr (\S\ref{sfdiag}).
Alternatively, these GORES might be the consequence of star formation
triggered by rapid flybys and rare ongoing direct (satellite-satellite)
mergers in clusters.


Finally, our statistics lead us to conclude that a fraction
$0.13\times(1-0.23)/0.18=0.56$ of the GORES within the virial cylinder are within
the virial sphere, so that \emph{44$\pm$2\% of GORES within the virial cylinder are outside
the virial sphere}, for all good-fitting Schemes.

Note that the asymptotic behaviour of the best-fitt models disagrees  
with the field value by at least 10\%. One may think that this  
discrepancy may be due to a different set of environment-dependent  
phenomenon become efficient in modulating galaxy properties at R$ > 2$\,\rv
 from the centres of clusters, or to the oversimplified models we used  
to deproject the fraction of GORES. However, according to the  
cosmological simulation we analysed here, the high absolute LOS  
velocity galaxies at $R>R_{\rm v}$ ought to lie at similar very far  
distances from the cluster centres (12 virial radii, see Fig.~\ref{rvsR}) as  
the field galaxies. Therefore, the discontinuity in the observed  
fraction of GORES between these galaxies and the field is puzzling and  
remains an open question.

\subsection{Epilogue}

In summary, our work shows that 
the galaxy properties in and around clusters are not 
simple functions of stellar mass and local environment
(e.g. \citealp{haines07,vonderlinden10}), but also 
of absolute LOS velocity. With these velocities, it is then statistically feasible to
 segregate the infalling and backsplash galaxies at any
 cluster-centric radius, including cluster outskirts. 
This velocity modulation  will be better seen in the properties of the
 relatively low-mass galaxies, because they are the first ones to be
 influenced by any changes in their immediate environment
(see \citealp{mahajand11} for such an effect seen in the nearby Coma cluster). 
It would be worthwhile to extend our analysis using  other indicators
of very recent star formation to 1) confirm our conclusion that the quenching of
star formation occurs in a single passage through the cluster, and 2) assess
the fraction of galaxies with very recent star formation among the virialised
population. The survival of low-mass galaxies and of star formation in galaxies
 of all mass, as they pass through the cluster core, lead to important issues that
 should also be addressed with high-resolution cosmological hydrodynamical simulations.

 \section*{Acknowledgements}

SM is supported by grants from ORSAS, UK, and the University
 of Birmingham. We thank the anonymous referee for useful comments and references.
GAM thanks G. Murante for providing the cosmological
 simulation, and A. Biviano, G. Kauffmann, M. Sarzi, T.~J. Ponman, and V. Wild
 for useful conversations. SM is grateful to B.~M. Poggianti for a useful discussion.
 Funding for the Sloan Digital Sky Survey (SDSS) has
 been provided by the Alfred P. Sloan Foundation, the Participating Institutions, the
 National Aeronautics and Space Administration, the National Science
 Foundation, the U.S. Department of Energy, the Japanese
 Monbukagakusho, and the Max Planck Society. The SDSS Web site is
 http://www.sdss.org/. 


\bibliography{mn3_v6}

\onecolumn

\appendix
\label{appendix}

\section[]{Kolmogorov-Smirnov Statistics}
In this appendix, we present tables of  
probabilities,
from the Kolmogorov-Smirnov 
test, that various
sub-samples of different physical quantities arise from the same parent
distributions (see \S~\ref{analysis}).

 \begin{table}
	\centering
 \caption{K-S test probabilities for the null hypothesis that pairs of
   subsamples have distributions of $M^*$ arising from the same parent
   distribution.
Columns $v$1, $v$2, and $v$3 correspond to  $|v_{\rm LOS}|/\sigma_v\!=\!0$--1,
1--2, and 2--3, respectively.
Statistically significant differences between subsample pairs are highlighted
 in bold.}
\tabcolsep 3pt
	\small{
		\begin{tabular}{|c|c|c|c|c|c|c|c|c|c|c|c|c|c|}
 \hline
     & \multicolumn{3}{|c|}{$R/R_{\rm v}\leq$0.5}  & \multicolumn{3}{|c|}{0.5$<$$R/R$$_{\rm v}\leq$1.0}  &\multicolumn{3}{|c|}{1.0$<$$R/R$$_{\rm v}\leq$1.5}   &\multicolumn{3}{|c|}{1.5$<$$R/R$$_{\rm v}\leq$2.0} &
    field   \\ \hline
     & $v$1 & $v$2 & $v$3 & $v$1 & $v$2 & $v$3 & $v$1 & $v$2 & $v$3 & $v$1 & $v$2 & $v$3 &        \\ \hline
   & 1 & 2 & 3 & 4 & 5 & 6 & 7 & 8 & 9 & 10 & 11 & 12 & \\

\hline
 1 & 1.0E+00 &  {\bf{7.3E-08}} &  {\bf{2.5E-09}} &  {\bf{4.2E-20}} &  {\bf{2.9E-20}} &  {\bf{1.6E-09}} &  {\bf{1.2E-12}} &  {\bf{1.8E-09}} &  {\bf{4.1E-08}} &  {\bf{2.7E-11}} &  {\bf{1.4E-10}} &  {\bf{8.5E-06}} &  {\bf{0.0E+00}} \\
 2 &  & 1.0E+00 &  {\bf{1.7E-02}} &  {\bf{3.1E-03}} &  {\bf{6.2E-06}} &  {\bf{9.9E-04}} &  {\bf{5.9E-03}} & 6.1E-02 &  {\bf{1.2E-03}} &  {\bf{2.1E-02}} & 5.5E-02 & 6.5E-02 &  {\bf{2.6E-13}} \\
 3 &  &  & 1.0E+00 & 1.7E-01 & 3.6E-01 & 3.6E-01 &  {\bf{4.3E-02}} & 1.2E-01 & 2.6E-01 &  {\bf{3.6E-02}} & 3.2E-01 & 7.3E-01 & 2.6E-01 \\
 4 & &  &  & 1.0E+00 &  {\bf{2.8E-02}} & 1.3E-01 & 2.1E-01 & 7.6E-01 & 2.3E-01 & 1.4E-01 & 8.8E-01 & 6.7E-01 &  {\bf{5.6E-07}} \\
 5 &  &  &  &  & 1.0E+00 & 6.5E-01 &  {\bf{2.5E-02}} & 8.8E-02 & 8.9E-01 &  {\bf{3.3E-03}} &  {\bf{3.2E-02}} & 6.4E-01 & 1.3E-01 \\
 6 &  &  & & &  & 1.0E+00 &  {\bf{3.8E-02}} & 9.9E-02 & 7.8E-01 &  {\bf{2.0E-02}} & 7.9E-02 & 2.6E-01 & 1.8E-01 \\
 7 & & &  &  & &  & 1.0E+00 & 7.9E-01 & 1.0E-01 & 9.7E-01 & 5.8E-01 & 6.7E-01 &  {\bf{2.6E-04}} \\
 8 &  &  &  &  &  &  & & 1.0E+00 & 1.5E-01 & 8.4E-01 & 7.9E-01 & 8.3E-01 &  {\bf{5.4E-03}} \\
 9 &  &  &  &  &  &  &  & & 1.0E+00 &  {\bf{4.3E-02}} & 9.9E-02 & 5.2E-01 & 4.3E-01 \\
 10 &  &  &  &  &  & &  &  &  & 1.0E+00 & 4.9E-01 & 7.0E-01 &  {\bf{5.4E-06}} \\
 11 &  &  & &  &  &  &  &  &  &  & 1.0E+00 & 9.1E-01 &  {\bf{3.7E-03}} \\
 12 & &  &  &  &  &  &  &  &  &  &  & 1.0E+00 & 3.8E-01 \\
  \hline
 \end{tabular}}
 \label{table:lgm}

 	\centering
 \caption{Same as Table~\ref{table:lgm}, but with BCGs excluded.}
 \label{table:lgm-nobcg}
\tabcolsep 3pt
	\small{
		\begin{tabular}{|c|c|c|c|c|c|c|c|c|c|c|c|c|c|}
 \hline
     & \multicolumn{3}{|c|}{$R/R_{\rm v}\leq$0.5}  & \multicolumn{3}{|c|}{0.5$<$$R/R$$_{\rm v}\leq$1.0}  &\multicolumn{3}{|c|}{1.0$<$$R/R$$_{\rm v}\leq$1.5}   &\multicolumn{3}{|c|}{1.5$<$$R/R$$_{\rm v}\leq$2.0} &
    field   \\ \hline
     & $v$1 & $v$2 & $v$3 & $v$1 & $v$2 & $v$3 & $v$1 & $v$2 & $v$3 & $v$1 & $v$2 & $v$3 &        \\ \hline
   & 1 & 2 & 3 & 4 & 5 & 6 & 7 & 8 & 9 & 10 & 11 & 12 & \\

\hline
 1 & 1.0E+00 &  {\bf{5.2E-08}} &  {\bf{6.6E-10}} &  {\bf{1.5E-20}} &  {\bf{1.2E-21}} &  {\bf{1.2E-09}} &  {\bf{6.9E-15}} &  {\bf{9.0E-10}} &  {\bf{6.5E-09}} &  {\bf{2.0E-12}} &  {\bf{9.5E-11}} &  {\bf{3.9E-06}} &  {\bf{0.0E+00}} \\
 2 &  & 1.0E+00 &  {\bf{9.2E-03}} &  {\bf{1.5E-03}} &  {\bf{1.7E-06}} &  {\bf{2.3E-04}} &  {\bf{8.9E-04}} &  {\bf{3.6E-02}} &  {\bf{5.1E-04}} &  {\bf{4.0E-03}} &  {\bf{3.8E-02}} &  {\bf{3.8E-02}} &  {\bf{3.9E-13}} \\
 3 &  &  & 1.0E+00 & 2.0E-01 & 4.3E-01 & 2.7E-01 & 9.8E-02 & 1.3E-01 & 2.5E-01 &  {\bf{4.4E-02}} & 2.2E-01 & 7.4E-01 & 3.4E-01 \\
 4 &  &  &  & 1.0E+00 &  {\bf{2.3E-02}} & 6.3E-02 & 5.5E-01 & 7.6E-01 & 1.7E-01 & 3.0E-01 & 8.5E-01 & 6.5E-01 &  {\bf{2.6E-06}} \\
 5 &  &  &  &  & 1.0E+00 & 6.9E-01 &  {\bf{4.6E-02}} & 1.0E-01 & 8.4E-01 &  {\bf{5.4E-03}} &  {\bf{2.5E-02}} & 6.7E-01 & 5.8E-01 \\
 6 &  &  &  &  &  & 1.0E+00 & 6.7E-02 & 7.7E-02 & 7.3E-01 &  {\bf{3.4E-02}} &  {\bf{4.1E-02}} & 2.6E-01 & 1.5E-01 \\
 7 &  &  &  &  &  &  & 1.0E+00 & 9.8E-01 & 1.0E-01 & 8.9E-01 & 8.1E-01 & 7.8E-01 &  {\bf{4.5E-03}} \\
 8 &  &  &  &  &  &  &  & 1.0E+00 & 1.3E-01 & 8.9E-01 & 7.3E-01 & 8.7E-01 &  {\bf{1.7E-02}} \\
 9 &  &  &  &  &  &  &  &  & 1.0E+00 & 5.2E-02 & 6.9E-02 & 4.5E-01 & 1.9E-01 \\
10 &  &  &  &  &  &  &  &  &  & 1.0E+00 & 4.9E-01 & 6.8E-01 &  {\bf{5.8E-05}} \\
11 &  &  &  &  &  &  &  &  &  &  & 1.0E+00 & 9.0E-01 &  {\bf{8.2E-03}} \\
12 &  &  &  &  &  &  &  &  &  &  & & 1.0E+00 & 5.3E-01 \\
 \end{tabular}}
\end{table}

\begin{table}
	\centering
 \caption{Same as Table~\ref{table:lgm}, but for the difference between
 the $(g-r)^{0.1}$ colour and the best-fitted red sequence (RS).}
 \label{table:gr}
\tabcolsep 3pt
	\small{
		\begin{tabular}{|c|c|c|c|c|c|c|c|c|c|c|c|c|c|}
 \hline
     & \multicolumn{3}{c|}{$R/R_{\rm v}\leq$0.5}  & \multicolumn{3}{c|}{0.5$<$$R/R$$_{\rm v}\leq$1.0}  &\multicolumn{3}{c|}{1.0$<$$R/R$$_{\rm v}\leq$1.5}   &\multicolumn{3}{c|}{1.5$<$$R/R$$_{\rm v}\leq$2.0}  & field   \\ \hline
     & $v$1 & $v$2 & $v$3 & $v$1 & $v$2 & $v$3 & $v$1 & $v$2 & $v$3 & $v$1 & $v$2 & $v$3 &        \\ \hline
   & 1 & 2 & 3 & 4 & 5 & 6 & 7 & 8 & 9 & 10 & 11 & 12 & \\

 \hline 
  1 & 1.0E+00 & 1.2E-01 &  {\bf{1.0E-04}} &  {\bf{5.9E-30}} &  {\bf{3.4E-29}} &  {\bf{1.3E-10}} &  {\bf{0.0E+00}} &  {\bf{0.0E+00}} &  {\bf{2.0E-29}} &  {\bf{0.0E+00}} &  {\bf{0.0E+00}} &  {\bf{9.0E-27}} &  {\bf{0.0E+00}} \\
  2 &  & 1.0E+00 &  {\bf{1.7E-02}} &  {\bf{1.5E-16}} &  {\bf{6.1E-17}} &  {\bf{1.8E-07}} &  {\bf{1.0E-41}} &  {\bf{7.3E-41}} &  {\bf{6.2E-23}} &  {\bf{1.4E-45}} &  {\bf{8.4E-36}} &  {\bf{7.6E-20}} &  {\bf{0.0E+00}} \\
  3 &  & & 1.0E+00 &  {\bf{5.8E-05}} &  {\bf{4.6E-06}} &  {\bf{1.6E-02}} &  {\bf{1.2E-13}} &  {\bf{4.5E-16}} &  {\bf{1.8E-12}} &  {\bf{3.9E-15}} &  {\bf{2.5E-13}} &  {\bf{2.4E-08}} &  {\bf{0.0E+00}} \\
  4 &  &  &  & 1.0E+00 & 5.6E-02 & 3.3E-01 &  {\bf{1.3E-11}} &  {\bf{1.2E-14}} &  {\bf{3.8E-10}} &  {\bf{2.2E-15}} &  {\bf{2.2E-11}} &  {\bf{1.1E-07}} &  {\bf{0.0E+00}} \\
  5 &  &  &  &  & 1.0E+00 & 8.1E-01 &  {\bf{1.1E-04}} &  {\bf{1.6E-07}} &  {\bf{1.3E-06}} &  {\bf{5.4E-06}} &  {\bf{1.8E-05}} &  {\bf{1.0E-04}} &  {\bf{0.0E+00}} \\
  6 &  &  &  &  &  & 1.0E+00 &  {\bf{2.4E-03}} &  {\bf{4.1E-05}} &  {\bf{6.1E-05}} &  {\bf{8.0E-04}} &  {\bf{9.9E-04}} &  {\bf{5.5E-03}} &  {\bf{4.9E-29}} \\
  7 &  &  &  &  & &  & 1.0E+00 & 1.1E-01 &  {\bf{1.8E-02}} & 6.1E-01 & 2.7E-01 & 1.0E-01 &  {\bf{0.0E+00}} \\
  8 &  &  &  &  &  &  & & 1.0E+00 & 3.3E-01 & 1.3E-01 & 6.5E-01 & 3.6E-01 &  {\bf{1.7E-23}} \\
  9 &  &  &  &  &  &  &  &  & 1.0E+00 &  {\bf{5.0E-02}} & 1.5E-01 & 1.5E-01 &  {\bf{3.0E-06}} \\
  10 &  &  &  &  &  &  &  &  &  & 1.0E+00 & 8.1E-01 & 4.8E-01 &  {\bf{0.0E+00}} \\
  11 &  &  &  &  &  &  &  &  &  &  & 1.0E+00 & 5.6E-01 &  {\bf{1.4E-31}} \\
  12 & &  &  &  &  &  &  &  &  &  &  & 1.0E+00 &  {\bf{1.9E-16}} \\
 \hline

 \end{tabular}}
   
	\centering
 \caption{Same as Table~\ref{table:lgm}, but for SFR/$M^*$.}
 \label{table:ssfr}
\tabcolsep 3pt
	\small{
		\begin{tabular}{|c|c|c|c|c|c|c|c|c|c|c|c|c|c|}
 \hline
     & \multicolumn{3}{c|}{$R/R_{\rm v}\leq$0.5}  & \multicolumn{3}{c|}{0.5$<$$R/R$$_{\rm v}\leq$1.0}  &\multicolumn{3}{c|}{1.0$<$$R/R$$_{\rm v}\leq$1.5}   &\multicolumn{3}{c|}{1.5$<$$R/R$$_{\rm v}\leq$2.0} & field   \\ \hline
     & $v$1 & $v$2 & $v$3 & $v$1 & $v$2 & $v$3 & $v$1 & $v$2 & $v$3 & $v$1 & $v$2 & $v$3 &        \\ \hline
   & 1 & 2 & 3 & 4 & 5 & 6 & 7 & 8 & 9 & 10 & 11 & 12 & \\

 \hline
  1 & 1.0E+00 &  {\bf{5.8E-04}} &  {\bf{1.0E-04}} &  {\bf{1.0E-26}} &  {\bf{8.1E-28}} &  {\bf{4.7E-12}} &  {\bf{0.0E+00}} &  {\bf{5.6E-45}} &  {\bf{1.1E-22}} &  {\bf{0.0E+00}} &  {\bf{1.8E-40}} &  {\bf{1.5E-21}} &  {\bf{0.0E+00}} \\
  2 &  & 1.0E+00 &  {\bf{1.7E-02}} &  {\bf{1.0E-08}} &  {\bf{6.2E-13}} &  {\bf{1.3E-06}} &  {\bf{4.1E-25}} &  {\bf{1.1E-25}} &  {\bf{1.2E-16}} &  {\bf{2.1E-33}} &  {\bf{1.7E-21}} &  {\bf{1.2E-14}} &  {\bf{0.0E+00}} \\
  3 &  &  & 1.0E+00 &  {\bf{3.4E-03}} &  {\bf{1.4E-02}} &  {\bf{3.1E-02}} &  {\bf{1.5E-06}} &  {\bf{1.1E-07}} &  {\bf{7.5E-08}} &  {\bf{1.0E-07}} &  {\bf{2.5E-06}} &  {\bf{3.4E-06}} &  {\bf{1.6E-21}} \\
  4 &  &  & & 1.0E+00 & 6.1E-02 & 8.4E-02 &  {\bf{4.1E-07}} &  {\bf{1.6E-09}} &  {\bf{4.5E-09}} &  {\bf{2.3E-11}} &  {\bf{3.2E-08}} &  {\bf{7.9E-07}} &  {\bf{0.0E+00}} \\
  5 &  & &  &  & 1.0E+00 & 4.7E-01 &  {\bf{6.2E-04}} &  {\bf{1.1E-04}} &  {\bf{6.1E-06}} &  {\bf{3.0E-05}} &  {\bf{3.4E-04}} &  {\bf{1.8E-04}} &  {\bf{1.4E-26}} \\
  6 &  & &  &  &  & 1.0E+00 &  {\bf{9.1E-03}} &  {\bf{2.3E-02}} &  {\bf{6.3E-03}} &  {\bf{4.1E-03}} &  {\bf{1.6E-02}} &  {\bf{1.1E-02}} &  {\bf{5.5E-07}} \\
  7 &  &  &  &  &  &  & 1.0E+00 & 1.5E-01 &  {\bf{2.5E-03}} & 3.0E-01 &  {\bf{7.3E-02}} &  {\bf{2.9E-02}} &  {\bf{1.9E-17}} \\
  8 &  &  &  &  &  &  &  & 1.0E+00 &  {\bf{9.7E-02}} & 4.6E-01 & 5.2E-01 & 4.5E-01 &  {\bf{2.9E-05}} \\
  9 &  &  &  &  & &  &  &  & 1.0E+00 &  {\bf{1.7E-02}} & 8.9E-02 & 7.5E-01 & 1.1E-01 \\
  10 &  &  &  &  & &  &  &  &  & 1.0E+00 & 4.3E-01 & 1.9E-01 &  {\bf{2.7E-14}} \\
  11 &  &  &  &  &  &  &  &  &  &  & 1.0E+00 & 6.0E-01 &  {\bf{1.8E-08}} \\
  12 &  & &  &  &  &  &  &  &  &  &  & 1.0E+00 &  {\bf{2.9E-03}} \\
 \hline

 \end{tabular}}

	\centering
 \caption{Same as Table~\ref{table:lgm}, but for H$\delta$.}
 \label{table:hd}
\tabcolsep 3pt
	\small{
		\begin{tabular}{|c|c|c|c|c|c|c|c|c|c|c|c|c|c|}
 \hline
     & \multicolumn{3}{c|}{$R/R_{\rm v}\leq$0.5}  & \multicolumn{3}{c|}{0.5$<$$R/R$$_{\rm v}\leq$1.0}  &\multicolumn{3}{c|}{1.0$<$$R/R$$_{\rm v}\leq$1.5}   &\multicolumn{3}{c|}{1.5$<$$R/R$$_{\rm v}\leq$2.0}  & field   \\ \hline
     & $v$1 & $v$2 & $v$3 & $v$1 & $v$2 & $v$3 & $v$1 & $v$2 & $v$3 & $v$1 & $v$2 & $v$3 &        \\ \hline
   & 1 & 2 & 3 & 4 & 5 & 6 & 7 & 8 & 9 & 10 & 11 & 12 & \\

 \hline
 1 & 1.0E+00 &  {\bf{1.7E-06}} &  {\bf{7.0E-09}} &  {\bf{0.0E+00}} &  {\bf{0.0E+00}} &  {\bf{6.1E-22}} &  {\bf{0.0E+00}} &  {\bf{0.0E+00}} &  {\bf{8.9E-37}} &  {\bf{0.0E+00}} &  {\bf{0.0E+00}} &  {\bf{1.1E-27}} &  {\bf{0.0E+00}} \\
 2 &  & 1.0E+00 &  {\bf{6.6E-04}} &  {\bf{1.4E-18}} &  {\bf{2.6E-24}} &  {\bf{9.6E-14}} &  {\bf{3.7E-38}} &  {\bf{7.7E-39}} &  {\bf{4.4E-25}} &  {\bf{1.5E-39}} &  {\bf{6.2E-35}} &  {\bf{1.6E-17}} &  {\bf{0.0E+00}} \\
 3 &  &  & 1.0E+00 &  {\bf{6.0E-03}} &  {\bf{8.5E-05}} &  {\bf{2.8E-03}} &  {\bf{7.3E-07}} &  {\bf{2.8E-10}} &  {\bf{2.0E-09}} &  {\bf{2.7E-06}} &  {\bf{9.4E-08}} &  {\bf{1.2E-04}} &  {\bf{4.4E-36}} \\
 4 &  &  &  & 1.0E+00 &  {\bf{2.4E-03}} &  {\bf{5.8E-03}} &  {\bf{7.2E-07}} &  {\bf{1.9E-11}} &  {\bf{4.4E-11}} &  {\bf{1.3E-08}} &  {\bf{3.6E-09}} & {\bf{1.4E-04}} &  {\bf{0.0E+00}} \\
 5 &  &  &  &  & 1.0E+00 & 7.8E-01 & 2.0E-01 &  {\bf{2.4E-04}} &  {\bf{2.7E-06}} & 5.5E-02 &  {\bf{1.3E-02}} & 7.8E-02 &  {\bf{1.6E-42}} \\
 6 &  &  &  &  & & 1.0E+00 & 2.1E-01 &  {\bf{1.6E-02}} &  {\bf{6.4E-04}} & 2.2E-01 & 7.7E-02 & 1.7E-01 &  {\bf{1.0E-12}} \\
 7 &  &  &  &  &  &  & 1.0E+00 &  {\bf{3.6E-02}} &  {\bf{3.6E-04}} & 1.0E-01 & 3.1E-01 & 5.7E-01 &  {\bf{6.1E-42}} \\
 8 &  &  &  &  &  &  &  & 1.0E+00 & 1.6E-01 &  {\bf{1.8E-02}} & 3.2E-01 & 4.3E-01 &  {\bf{5.4E-13}} \\
 9 &  &  &  &  &  &  &  &  & 1.0E+00 &  {\bf{9.9E-04}} &  {\bf{1.0E-02}} & 6.6E-02 &  {\bf{1.3E-02}} \\
 10 &  &  &  & &  &  &  &  & & 1.0E+00 & 4.1E-01 & 8.6E-01 &  {\bf{1.0E-43}} \\
 11 &  &  &  &  &  &  &  &  &  & & 1.0E+00 & 9.8E-01 &  {\bf{9.1E-19}} \\
 12 &  &  &  &  &  &  &  &  &  &  &  & 1.0E+00 &  {\bf{7.3E-09}} \\
 \hline

 \end{tabular}}
 \end{table} 

\begin{table}
	\centering
 \caption{Same as Table~\ref{table:lgm}, but for  D$_n$4000.}
 \label{table:d4}
\tabcolsep 3pt
	\small{
		\begin{tabular}{|c|c|c|c|c|c|c|c|c|c|c|c|c|c|}
 \hline
     & \multicolumn{3}{c|}{$R/R_{\rm v}\leq$0.5}  & \multicolumn{3}{c|}{0.5$<$$R/R$$_{\rm v}\leq$1.0}  &\multicolumn{3}{c|}{1.0$<$$R/R$$_{\rm v}\leq$1.5}   &\multicolumn{3}{c|}{1.5$<$$R/R$$_{\rm v}\leq$2.0} & field    \\ \hline
     & $v$1 & $v$2 & $v$3 & $v$1 & $v$2 & $v$3 & $v$1 & $v$2 & $v$3 & $v$1 & $v$2 & $v$3 &        \\ \hline
   & 1 & 2 & 3 & 4 & 5 & 6 & 7 & 8 & 9 & 10 & 11 & 12 & \\

 \hline
 1 & 1.0E+00 &  {\bf{3.5E-09}} &  {\bf{5.1E-09}} &  {\bf{0.0E+00}} &  {\bf{0.0E+00}} &  {\bf{9.8E-15}} &  {\bf{0.0E+00}} &  {\bf{0.0E+00}} &  {\bf{8.8E-35}} &  {\bf{0.0E+00}} &  {\bf{0.0E+00}} &  {\bf{1.2E-27}} &  {\bf{0.0E+00}} \\
 2 &  & 1.0E+00 &  {\bf{1.1E-04}} &  {\bf{3.1E-23}} & {\bf{1.9E-23}} &  {\bf{8.4E-09}} &  {\bf{4.7E-40}} &  {\bf{6.2E-38}} &  {\bf{1.9E-24}} &  {\bf{1.4E-45}} &  {\bf{1.3E-35}} &  {\bf{5.2E-18}} &  {\bf{0.0E+00}} \\
 3 &  &  & 1.0E+00 &  {\bf{8.1E-03}} &  {\bf{7.2E-05}} &  {\bf{4.7E-02}} &  {\bf{7.4E-07}} &  {\bf{9.2E-09}} &  {\bf{5.7E-09}} &  {\bf{9.0E-07}} &  {\bf{3.7E-08}} &  {\bf{8.2E-05}} &  {\bf{1.2E-34}} \\
 4 &  &  &  & 1.0E+00 &  {\bf{2.2E-02}} & 1.0E-01 &  {\bf{1.2E-06}} &  {\bf{4.8E-09}} &  {\bf{2.5E-07}} &  {\bf{1.6E-11}} &  {\bf{5.6E-09}} &  {\bf{1.9E-05}} &  {\bf{0.0E+00}} \\
 5 &  &  & & & 1.0E+00 & 2.8E-01 & 6.3E-02 &  {\bf{8.2E-04}} &  {\bf{9.8E-05}} &  {\bf{1.0E-03}} &  {\bf{5.8E-03}} &  {\bf{1.7E-02}} &  {\bf{0.0E+00}} \\
 6 &  &  &  &  &  & 1.0E+00 &  {\bf{1.4E-02}} &  {\bf{5.6E-04}} &  {\bf{5.1E-05}} &  {\bf{1.7E-02}} &  {\bf{5.2E-03}} & 9.5E-02 &  {\bf{1.1E-17}} \\
 7 &  &  &  & &  &  & 1.0E+00 & 7.1E-02 &  {\bf{7.5E-03}} & 1.1E-01 & 6.9E-02 & 1.8E-01 &  {\bf{0.0E+00}} \\
 8 & &  & &  &  &  &  & 1.0E+00 & 2.1E-01 & 1.2E-01 & 9.4E-01 & 1.0E-01 &  {\bf{7.4E-17}} \\
 9 &  &  &  & & &  &  & & 1.0E+00 &  {\bf{2.1E-02}} & 1.1E-01 &  {\bf{4.9E-02}} &  {\bf{1.3E-03}} \\
 10 &  &  &  &  & &  &  &  &  & 1.0E+00 & 3.5E-01 & 5.9E-01 &  {\bf{2.5E-38}} \\
 11 &  & &  &  &  &  &  &  &  &  & 1.0E+00 & 1.2E-01 &  {\bf{1.5E-21}} \\
 12 &  &  &  &  &  &  &  &  &  &  &  & 1.0E+00 &  {\bf{4.1E-10}} \\
 \hline
 \end{tabular}}
 \end{table}

 \label{lastpage}

\end{document}